\documentclass[12pt,a4paper]{article}  
\usepackage{psfrag}   
\usepackage{graphics}  
\usepackage{amssymb,epsfig,amsmath,euscript,array} 
\usepackage{cite}  
\makeatletter
\@addtoreset{equation}{section}
\makeatother
\renewcommand{\theequation}{\thesection.\arabic{equation}}

\newcommand{\startappendix}{
\setcounter{section}{0}
\renewcommand{\thesection}{\Alph{section}}
\renewcommand{\theequation}{\Alph{section}.\arabic{equation}}}

\newcommand{\Appendix}[1]{
\refstepcounter{section}
\begin{flushleft}
{\Large\bf Appendix \thesection: #1}
\end{flushleft}}

\newcounter{multieqs}

\newcommand{\be}{\begin{equation}}  
\newcommand{\ee}{\end{equation}}

\newcommand{\bra}[1]{\langle #1|}  
\newcommand{\ket}[1]{|#1 \rangle}

\newcommand{\bm}[1]{\mbox{\boldmath $#1$}}

\def\bd{\begin{document}}  
\def\ed{\end{document}}  
\def\nn{\nonumber}  
\def\bea{\begin{eqnarray}}  
\def\eea{\end{eqnarray}}  
\let\bm=\bibitem  
\let\la=\label  
  
\def\npb#1#2#3{Nucl. Phys. {\bf{B#1}} #3 (#2)}  
\def\plb#1#2#3{Phys. Lett. {\bf{#1B}} #3 (#2)}  
\def\prl#1#2#3{Phys. Rev. Lett. {\bf{#1}} #3 (#2)}  
\def\prd#1#2#3{Phys. Rev. {D \bf{#1}} #3 (#2)}  
\def\cmp#1#2#3{Comm. Math. Phys. {\bf{#1}} #3 (#2)}  
\def\cqg#1#2#3{Class. Quantum Grav. {\bf{#1}} #3 (#2)}  
\def\nppsa#1#2#3{Nucl. Phys. B (Proc. Suppl.) {\bf{#1A}}#3 (#2)}  
\def\ap#1#2#3{Ann. of Phys. {\bf{#1}} #3 (#2)}  
\def\ijmp#1#2#3{Int. J. Mod. Phys. {\bf{A#1}} #3 (#2)}  
\def\rmp#1#2#3{Rev. Mod. Phys. {\bf{#1}} #3 (#2)}  
\def\mpla#1#2#3{Mod. Phys. Lett. {\bf A#1} #3 (#2)}  
\def\jhep#1#2#3{J. High Energy Phys. {\bf #1} #3 (#2)}  
\def\atmp#1#2#3{Adv. Theor. Math. Phys. {\bf #1} #3 (#2)}  
  
\newcommand{\EQ}[1]{\begin{equation} #1 \end{equation}}  
\newcommand{\AL}[1]{\begin{subequations}\begin{align} #1 \end{align}\end{subequations}}  
\newcommand{\SP}[1]{\begin{equation}\begin{split} #1 \end{split}\end{equation}}  
\newcommand{\ALAT}[2]{\begin{subequations}\begin{alignat}{#1} #2 \end{alignat}\end{subequations}}  
\def\beqa{\begin{eqnarray}}   
\def\eeqa{\end{eqnarray}}   
\def\beq{\begin{equation}}   
\def\eeq{\end{equation}}   
  
\def\N{{\cal N}}  
\def\sst{\scriptscriptstyle}  
\def\thetabar{\bar\theta}  
\def\Tr{{\rm Tr}}  
\def\one{\mbox{1 \kern-.59em {\rm l}}}  
 \def\Nh{\hat{N}} 
  
\def\a{\alpha}      \def\da{{\dot\alpha}}  
\def\b{\beta}       \def\db{{\dot\beta}}  
\def\c{\gamma}  \def\G{\Gamma}  \def\cdt{\dot\gamma}  
\def\d{\delta}  \def\D{\Delta}  \def\ddt{\dot\delta}  
\def\e{\epsilon}        \def\vare{\varepsilon}  
\def\f{\phi}    \def\F{\Phi}    \def\vvf{\f}  
\def\h{\eta}  
\def\k{\kappa}  
\def\l{\lambda} \def\L{\Lambda}  
\def\m{\mu} \def\n{\nu}  
\def\o{\omega}  
\def\p{\pi} \def\P{\Pi}  
\def\r{\rho}  
\def\s{\sigma}  \def\S{\Sigma}  
\def\t{\tau}  
\def\th{\theta} \def\Th{\Theta} \def\vth{\vartheta}  
\def\X{\Xeta}  
\def\z{\zeta}  
  
\def\cA{{\cal A}} \def\cB{{\cal B}} \def\cC{{\cal C}}  
\def\cD{{\cal D}} \def\cE{{\cal E}} \def\cF{{\cal F}}  
\def\cG{{\cal G}} \def\cH{{\cal H}} \def\cI{{\cal I}}  
\def\cJ{{\cal J}} \def\cK{{\cal K}} \def\cL{{\cal L}}  
\def\cM{{\cal M}} \def\cN{{\cal N}} \def\cO{{\cal O}}  
\def\cP{{\cal P}} \def\cQ{{\cal Q}} \def\cR{{\cal R}}  
\def\cS{{\cal S}} \def\cT{{\cal T}} \def\cU{{\cal U}}  
\def\cV{{\cal V}} \def\cW{{\cal W}} \def\cX{{\cal X}}  
\def\cY{{\cal Y}} \def\cZ{{\cal Z}}

\def\ua{\underline{\alpha}}  
\def\ub{\underline{\phantom{\alpha}}\!\!\!\beta}  
\def\uc{\underline{\phantom{\alpha}}\!\!\!\gamma}  
\def\um{\underline{\mu}}  
\def\ud{\underline\delta}  
\def\ue{\underline\epsilon}  
\def\una{\underline a}\def\unA{\underline A}  
\def\unb{\underline b}\def\unB{\underline B}  
\def\unc{\underline c}\def\unC{\underline C}  
\def\und{\underline d}\def\unD{\underline D}  
\def\une{\underline e}\def\unE{\underline E}  
\def\unf{\underline{\phantom{e}}\!\!\!\! f}\def\unF{\underline F}  
\def\unm{\underline m}\def\unM{\underline M}  
\def\unn{\underline n}\def\unN{\underline N}  
\def\unp{\underline{\phantom{a}}\!\!\! p}\def\unP{\underline P}  
\def\unq{\underline{\phantom{a}}\!\!\! q}  
\def\unQ{\underline{\phantom{A}}\!\!\!\! Q}  
\def\unH{\underline{H}}  
  
\def\As {{A \hspace{-6.4pt} \slash}\;}  
\def\bs {{b \hspace{-6.4pt} \slash}\;}  
\def\Ds {{D \hspace{-6.4pt} \slash}\;}  
\def\ds {{\del \hspace{-6.4pt} \slash}\;}  
\def\ss {{\s \hspace{-6.4pt} \slash}\;}  
\def\ks {{ k \hspace{-6.4pt} \slash}\;}  
\def\ps {{p \hspace{-6.4pt} \slash}\;}  
\def\pas {{{p_1} \hspace{-6.4pt} \slash}\;}  
\def\pbs {{{p_2} \hspace{-6.4pt} \slash}\;}  
  
\def\Fh{\hat{F}}  
\def\Vh{\hat{V}}  
\def\Xh{\hat{X}}  
\def\ah{\hat{a}}  
\def\xh{\hat{x}}  
\def\yh{\hat{y}}  
\def\ph{\hat{p}}  
\def\xih{\hat{\xi}}  
  
\def\psit{\tilde{\psi}}  
\def\Psit{\tilde{\Psi}}  
\def\tht{\tilde{\th}}  
   
\def\At{\tilde{A}}  
\def\Qt{\tilde{Q}}  
\def\Rt{\tilde{R}}  
\def\Nt{\tilde{N}}  
  
\def\at{\tilde{a}}  
\def\st{\tilde{s}}  
\def\ft{\tilde{f}}  
\def\pt{\tilde{p}}  
\def\qt{\tilde{q}}  
\def\vt{\tilde{v}}  
\def\nt{\tilde{n}}  
  
\def\delb{\bar{\partial}}  
\def\bz{\bar{z}}  
\def\bD{\bar{D}}  
\def\bB{\bar{B}}  
  
\def\bk{{\bf k}}  
\def\bl{{\bf l}}  
\def\bp{{\bf p}}  
\def\bq{{\bf q}}  
\def\br{{\bf r}}  
\def\bx{{\bf x}}  
\def\by{{\bf y}}  
\def\bR{{\bf R}}  
\def\bV{{\bf V}}  
  
\def\d{\delta}\def\D{\Delta}\def\ddt{\dot\delta}  
  
\def\pa{\partial} \def\del{\partial}  
\def\xx{\times}  
\def\uno{\mbox{1 \kern-.59em {\rm l}}}    
  
\def\trp{^{\top}}  
\def\inv{^{-1}}  
\def\dag{{^{\dagger}}}  
\def\pr{^{\prime}}  
  
\def\rar{\rightarrow}  
\def\lar{\leftarrow}  
\def\lrar{\leftrightarrow}  
  
\newcommand{\0}{\,\!}      %this is just NOTHING!  
\def\one{1\!\!1\,\,}  
\def\im{\imath}  
\def\jm{\jmath}  
  
\newcommand{\tr}{\mbox{tr}}  
\newcommand{\slsh}[1]{/ \!\!\!\! #1}  
  
\def\vac{|0\rangle}  
\def\lvac{\langle 0|}  
  
\def\hlf{\frac{1}{2}}  
\def\ove#1{\frac{1}{#1}}  

\def\Box{\square}  
\def\ZZ{\mathbb{Z}}  
\def\CC#1{({\bf #1})}  
\def\bcomment#1{}  
\def\bfhat#1{{\bf \hat{#1}}}  
\def\VEV#1{\left\langle #1\right\rangle}  

\newcommand{\ex}[1]{{\rm e}^{#1}} \def\ii{{\rm i}}  

\def\rr{{\rm r}} \def\rs{{\rm s}}\def\rv{{\rm v}}
\def\ri{{\rm i}}\def\rj{{\rm j}}
  
\newcommand{\lrbrk}[1]{\left(#1\right)}
\newcommand{\sfrac}[2]{{\textstyle\frac{#1}{#2}}}

\font\mybb=msbm10 at 12pt
\def\bb#1{\hbox{\mybb#1}}

\font\myBB=msbm10 at 18pt
\def\BB#1{\hbox{\myBB#1}}

\begin{document}  
  
\hfill{ hep-th/0306234}  
   
\vspace{20pt}  
   
\begin{center}  
  
{\Large \bf New tests of the pp-wave correspondence\\}   
%\vspace{10pt}  
%{\Large \bf  }  
%g new font for Z_2  
\vspace{30pt}  
   
{\bf George Georgiou,  
Valentin V.~Khoze and Gabriele Travaglini }

{\small \em
Centre for Particle Theory,
Department of Physics and IPPP,\\
University of Durham, Durham, DH1 3LE, UK
}

\vspace{10pt}  
  
{\sffamily \tt george.georgiou, valya.khoze,  
gabriele.travaglini@durham.ac.uk }

\vspace{30pt}  
{\bf Abstract}  
  
\end{center}  
The pp-wave/SYM  correspondence is an equivalence relation,
$H_{\rm string}= \Delta -J$, between the Hamiltonian
$H_{\rm string}$  of string field theory in the pp-wave background 
and the dilatation operator $\Delta $ in $\cN=4$ Super Yang-Mills
in the double scaling limit. We calculate matrix elements of 
these operators in string field theory and in gauge theory.
In the string theory Hilbert space we use the natural string 
basis, and in the gauge theory we use  the basis which is isomorphic to it.
States in this basis are specific linear 
combinations of the original BMN operators, and were constructed 
previously for the case of two scalar impurities. We extend this 
construction to incorporate BMN operators with vector and mixed impurities.
This enables us to verify from the 
gauge theory perspective two key properties of the three-string interaction 
vertex of Spradlin and Volovich:\\
{\bf (1)} the vanishing of the three-string amplitude 
for string states with one vector and one scalar impurity; 
and {\bf (2)} the relative minus sign in 
the string amplitude involving states with 
two vector impurities compared to that 
with two scalar impurities. 
This implies a spontaneous breaking of the 
$\bb{Z}_2$ symmetry of the  string field theory in the
pp-wave background.
Furthermore, we calculate the gauge theory matrix elements of 
$\Delta -J$ for states with an arbitrary number of scalar 
impurities. In all cases we find perfect agreement with 
the corresponding string amplitudes 
derived from the three-string vertex.

\vspace{0.5cm}  
  
\setcounter{page}{0}  
\thispagestyle{empty}  
\newpage

%%%%%%%%%%%%%% ordinary document (end) %%%%%%%%%%%%%%%%%%%%%%%%%%%%%%%%

\section{Introduction}
The pp-wave/SYM correspondence in its original form \cite{BMN} 
emphasises a duality relation
between the masses of string states and the 
anomalous dimensions \cite{zanon}
of the BMN operators in the dual gauge 
theory. 
This correspondence was  
further discussed and extended
in \cite{ver,gross2}, 
where it was expressed in the form \cite{gross2}
\be \label{ham}
{1\over \mu} \, {H}_{\rm string} = \Delta - J \ .
 \ee
Here ${H}_{\rm string}$ is the full string field 
theory Hamiltonian,
$\mu$ is the scale parameter of the pp-wave background, and $\Delta-J$ is 
the gauge theory dilatation operator (which in 
the radial quantisation formalism plays the r{\^o}le of the Hamiltonian 
${H}_{\rm SYM}$) minus the R-charge.  
The relation \eqref{ham} is expected to be exact and hold 
in the double scaling limit $N \sim J^2 \to \infty$
to all orders in the two free parameters of the theory, $g_2$ and $\l'$
\beqa 
\label{lampr}
\l' &=& \frac{g_{\rm YM}^2 N}{J^2} = \frac{1}{(\mu p^+ \a')^2}\ ,
\\ \cr
\label{gtwo}
g_2 &=& \frac{J^2}{N}= 4 \pi \,g_{\rm st}\, (\mu p^+ \a')^2 \ .
\eeqa 
On the gauge theory side, $\l'$ is the effective coupling 
constant of the BMN sector, and $g_2$ is the effective genus 
counting parameter of Feynman diagrams
\cite{BMN,KPSS,Constable1}. 
The right hand sides of \eqref{lampr}, \eqref{gtwo} express $\l'$ and
$g_2$ in terms
of the pp-wave string theory parameters to the effect that 
$1/\l' \propto \mu$ measures the deviation from the flat
background and $g_2$ is proportional to the string coupling 
$g_{\rm st}$ in IIB.

Since the two Hamiltonians, ${H}_{\rm string}$ and $\Delta$, act
on the states of two different theories, the duality 
relation \eqref{ham} requires an isomorphism between 
the Hilbert spaces of the light-cone gauge
pp-wave string field theory 
and of the BMN sector of the $\cN=4$ gauge theory. 
More specifically, we need to establish a one-to-one correspondence 
between the bases of two 
theories, $\{ \ket{s_\alpha}^{\rm string} \}$ and 
$\{ \ket{s_\alpha}^{\rm SYM} \}$,
\be
\ket{s_\alpha}^{\rm string}\, \leftrightarrow \,
\ket{s_\alpha}^{\rm SYM} \ ,
\ee
which preserves the scalar product,
\be
{}^{\rm string}\langle s_\alpha \ket{s_\beta}^{\rm string} \,=\, 
{}^{\rm SYM}\langle s_\alpha \ket{s_\beta}^{\rm SYM} \ .
\ee
Then the correspondence \eqref{ham} holds at the matrix elements level,
\be 
\label{ham2}
{}^{\rm string}\bra{s_\alpha}\mu^{-1}{H}_{\rm string}
\ket{s_\beta}^{\rm string}\, =\,
{}^{\rm SYM}\bra{s_\alpha} \Delta - J \ket{s_\beta}^{\rm SYM}\ .
 \ee

The string field theory Hilbert space 
is equipped with a natural basis of multi-string states,
\be \label{nat}
\{ \ket{s_\alpha}^{\rm string} \}= \ket{{\rm string}_a}\ , \
\ket{{\rm string}_b}\otimes \ket{{\rm string}_c}\ , \
\ket{{\rm string}_d}\otimes \ket{{\rm string}_e}
\otimes \ket{{\rm string}_f}\ ,
\ldots
\ee
which diagonalises the free string Hamiltonian and
is automatically orthonormal.
Here $a,b,\ldots $ are the labels of single-string states.
This basis does not diagonalise the full string Hamiltonian, 
$H_{\rm string}$, since free string states in \eqref{nat} 
can interact (split and join). The splitting and joining of a single
string state is described by the three-string interaction,
and the corresponding matrix element on the left hand side of
\eqref{ham2} is
\be \label{hamst}
\bra{{\rm string}_a}{H}_{\rm string}^{\rm int} 
\ket{{\rm string}_b}\otimes \ket{{\rm string}_c} \ \equiv \ 
\bra{{\rm string}_a} 
\bra{{\rm string}_b} \bra{{\rm string}_c} H_3 \rangle\ .
\ee
Here the $\ket{H_3}$ is the three-string interaction vertex
in the light-cone string field theory in the pp-wave background.
The expression for this vertex was originally 
obtained by Spradlin and
Volovich \cite{SV1,SV2}
and further  studied and clarified by 
Pankiewicz and Stefanski  in 
\cite{AP,PS} and in other papers including 
\cite{CKPRT,Panknew}. 
Its expression 
is recalled in Appendix A.%
\footnote{For notational simplicity and in order 
to distinguish  this vertex from other proposals, we will sometimes
refer to the vertex of \cite{SV1,SV2,AP,PS} simply 
as the Spradlin-Volovich vertex.}
However,
there is a puzzle related to the three-string amplitudes
\eqref{hamst} built on the Spradlin-Volovich vertex 
which we would like to clarify in this paper, among other things.
The presence of a non-trivial
R--R field in the pp-wave background breaks the light-cone
Lorentz symmetry $SO(8)$ down to $SO(4)\times SO(4)
\times \bb{Z}_2$. The
two $SO(4)$'s rotate the first and the last four directions among
themselves, while the $\bb{Z}_2$ 
symmetry swaps these two groups of four directions. Apparently,
the $\bb{Z}_2$ part of the bosonic symmetry of the pp-wave
background is not respected by the Spradlin-Volovich
three-string interactions 
%gg
\cite{KKLP,LMP,SV2,CKPRT,KKPR}:
there is a relative minus sign in the string amplitude 
involving states with two oscillators along the first $SO(4)$
compared to that with two oscillators along the second $SO(4)$.
An unbroken $\bb{Z}_2$-invariance would not allow this to happen.
We will argue now that this minus sign 
implies a spontaneous breaking of the 
$\bb{Z}_2$ symmetry of the  string field theory in the pp-wave background.

The  ket-vertex $\ket{H_3}$ \eqref{139}, \eqref{Vb} of 
\cite{SV1,SV2} 
is built on the string state $\ket{0}$ which is the ground state
of the theory in 
flat background, but
not in the pp-wave background. At the same time, the external
string bra-states in \eqref{hamst} are built on the true 
pp-wave ground state $\ket{v}$. It was explained in \cite{CKPRT}
that these two states, $\ket{0}$ and $\ket{v}$, have an opposite
$\bb{Z}_2$ parity, i.e.~cannot be both invariant 
under $\bb{Z}_2$. Hence, it follows immediately \cite{SV2,CKPRT}
that the amplitude \eqref{hamst} 
is not invariant under the action of $\bb{Z}_2$, but
changes sign. 
In the recent paper \cite{Panknew}, 
the result of \cite{SV1,SV2}, which utilised the vacuum $\ket{0}$, 
was compared with an alternative formalism of constructing
$\ket{H_3}$ starting directly from the true ground-state $\ket{v}$. 
The two formalisms were found to be identical. 
Following \cite{Panknew,SV1,SV2} we choose the 
$\bb{Z}_2$-parity prescription
\be 
\bb{Z}_2 : \, \ket{0} \rightarrow \ket{0} \ , \qquad
\bb{Z}_2 : \, \ket{v} \rightarrow - \ket{v} \ .
\ee
This means that the vertex $\ket{H_3}$ built on $\ket{0}$ is
invariant under $\bb{Z}_2$, 
%g
%
%but the string ground-state,
%$\bra{v}$, and, hence, the external states,
%$\bra{{\rm string}_a} 
%\bra{{\rm string}_b} \bra{{\rm string}_c}$ in \eqref{hamst} 
%acquire a minus sign.
but the pp-wave string ground-state
$\bra{v}$ and, hence, the external states
$\bra{{\rm string}_a} 
\bra{{\rm string}_b} \bra{{\rm string}_c}$ in \eqref{hamst}, 
acquire a minus sign.
This implies a spontaneous breaking of the $\bb{Z}_2$ symmetry 
of the  string field theory in the pp-wave background, 
which is the physical reason for 
the minus sign of the matrix element discussed above.

One of the objectives of this paper is to verify 
with an independent gauge theory calculation this 
important
minus sign (and hence the spontaneous breaking of $\bb{Z}_2$), 
as well as the related fact that the three-string amplitude \eqref{hamst}
vanishes for string states with one direction along the first, 
and one one direction along the second $SO(4)$, 
i.e.~one vector and one scalar impurity in the gauge theory language.%
%\footnote{In field theory, the BMN operators 
%that are dual to string excitations
%in the first four directions carry impurities of the form 
%$D_\m Z$ (vector impurities).}

As already mentioned, and following \cite{gross2,bits3,Gomis},
in order to compare \eqref{hamst}
with matrix elements 
of the dilatation operator
in gauge theory via \eqref{ham2},
it is important to identify a basis in gauge theory 
which is isomorphic to the natural string basis \eqref{nat}.
We discuss this issue in section 2. 
States in the isomorphic to string 
basis, $\{ \ket{s_\alpha}^{\rm SYM} \}$,
are obtained from linear 
combinations of the original multi-trace BMN \cite{BMN} 
operators $\cO_\alpha (x)$, 
\be \label{isost}
\ket{s_\alpha}^{\rm SYM} \ = \ U_{\alpha \beta}\,\cO_\beta
(x=0) \ket{0} \ ,  
\ee
where $U_{\alpha \beta}$ is an $x$-independent matrix.
This matrix was determined in \cite{gross2,Gomis}
by requiring that \eqref{ham2} holds,
i.e.~that the 
known three-string interaction vertex of the pp-wave 
light-cone string field theory  
\cite{SV1,SV2} is reproduced from gauge theory matrix elements 
of the dilatation operator
involving BMN states (operators) with two scalar impurities. 

In this paper we will take $U_{\alpha \beta}$ determined in 
\cite{Gomis}, and use it to construct the gauge theory
basis \eqref{isost} for an arbitrary number of scalar  
impurities. 
With this in hand we can compute generic gauge theory matrix elements 
on the right hand side of \eqref{ham2}. 
The contributions on the left hand side of \eqref{ham2} 
are then computed using \eqref{hamst}.
We will verify \eqref{ham2} and hence 
the Spradlin-Volovich expression for $\ket{H_3}$ 
for 
{\it generic }bosonic impurities.
First successful steps in this direction have been already taken
in \cite{Gomis,Gomis2} at the level of arbitrary number of 
identical  scalar impurities.%
\footnote{For further tests of the correspondence in the 
open-closed string sector, see \cite{Gomis3}.}
However, 
the inclusion of vector impurities is essential 
in order to address in the gauge theory the 
two important properties of the three-string interaction 
discussed earlier:\\
{\bf (1) } the vanishing of the three-string amplitude 
for string states with one vector and one scalar impurity; and \\
{\bf (2) } the relative minus sign in the string amplitude involving 
states with two vector impurities 
compared to that with two scalar impurities.

In section \ref{2imp-sec}
we will verify {\bf (1)} and {\bf (2) } working at the two-impurity level, and 
will consider all representations of BMN operators
with two vector or scalar impurities,  
i.e.~symmetric traceless, antisymmetric and singlet.
By considering BMN operators with 
vector,  scalar and mixed (scalar+vector) impurities 
we explore and verify the correspondence \eqref{ham2}
for string states in all the  
directions of the two $SO(4)$ groups. 

In section \ref{anyimp-sec},  we will 
calculate the gauge theory matrix elements of 
$\Delta -J$ for states with an arbitrary number of scalar
impurities. 
Next we compute the corresponding three-string amplitudes 
derived from the three-string vertex 
and compare them to the field theory result, 
finding perfect agreement.

Finally, section 4 and section 6 are dedicated to 
computations of three-point correlators of BMN operators. 
These results are  used earlier in section 3 and 5
for the calculation of matrix elements. 
More specifically, 
in section 4 we compute the coefficient of the conformal 
three-point function of BMN operators with mixed 
(one scalar and one vector) impurities. 
In section 6 we generalise
this analysis 
to the case of 
BMN operators with an arbitrary number of scalar impurities.

%%%%%%%%%%%%%%%%%%%%% END INTRODUCTION %%%%%%%%%%%%%%%%%%%%%%%%

\section{The dilatation operator in SYM and the natural string basis}  
As mentioned earlier, the BMN basis in SYM which 
is isomorphic to the natural string basis in dual string field theory, 
is a certain linear combination \eqref{isost} 
of the original  BMN operators
$\cO_\alpha (x)$ proposed in \cite{BMN}. 
The states in the natural string basis are not identically equal 
to the original BMN operators since 
the former are automatically orthonormal, 
while the latter are not, and their overlaps 
depend on the string coupling $g_2$. In other words,
the matrix $U$ in \eqref{isost} is not simply the identity matrix.

Apart from the original BMN basis, 
there is another distinguished basis  
of the conformal primary BMN 
operators $\cO_{\D_\alpha} (x)$
which are the eigenstates of the dilatation operator 
$\D$ in gauge theory. This $\Delta$-BMN basis is again a linear
combination of the states from
the original BMN basis $\cO_\alpha (x)$ with a different 
$x$-independent matrix $U$. For BMN operators with scalar
impurities, this basis was constructed in \cite{BKPSS} and 
extended to include vector and mixed impurities in \cite{CKTvec}.
The  $\Delta$-BMN basis is particularly convenient since the 
two- and three-point correlation functions of $\Delta$-BMN
operators can be written in the simple canonical
form with a universal $x$-dependence, 
guaranteed by conformal invariance of the theory. 
For conformal primary operators with scalar impurities these
canonical correlators are particularly simple and are given by
\begin{equation} 
\label{2pt}
\langle {\cO}^{\dagger}_{\D_\alpha} (x) \cO_{\D_\beta} (0)
\rangle = 
\frac{\d_{\alpha \beta}}{(x^2)^{\Delta_\alpha}} \ ,
\end{equation}
\begin{equation} 
\label{3pt}
\langle \cO_{\Delta_1}(x_1) \cO_{\Delta_2}(x_2) 
\cO^{\dagger}_{\Delta_3}(x_3) \rangle  =
\frac{C_{123} }
{(x_{12}^2)^{\frac{\D_1+\D_2 -\D_3}{2}}
 (x_{13}^2)^{\frac{\D_1+\D_3 -\D_2}{2}}
 (x_{23}^2)^{\frac{\D_2+\D_3 -\D_1}{2}}} \ .
\end{equation}
Canonical expression for the correlators involving conformal
primary operators with vector impurities appear to be much less
illuminating and harder to interpret, however it was noted 
in \cite{CKTvec} that this difficulty is avoided and the 
correlators for all types of impurities can be expressed in the
same form, similar to \eqref{2pt} and \eqref{3pt}, if 
on the left hand sides of \eqref{2pt} and \eqref{3pt} we 
use a different notion of conjugation 
$\bar{\cO}$ instead of $\cO^\dagger$ \cite{CKTvec}. 
This  different  notion of operator conjugation is  
defined as  {\it hermitian conjugation} 
followed by an  {\it inversion} of the operator
argument $x'_{\mu} = x_{\mu} / x^2$.
Under inversion a scalar operator $\cO_{\Delta}(x)$
of conformal dimension $\Delta$ transforms as 
\beq
\label{scalinv}
\cO^\dagger_{\Delta} (x) \rightarrow 
\cO_{\Delta}^{\dagger'} (x') = 
x^{2 \Delta}  \cO^\dagger_{\Delta}(x) \ \ , 
\ \ x_{\m}\to x'_{\m} = {x_{\m} \over x^2}
\ ,
\eeq
while a vector or tensor operator (i.e.~operator with vector 
impurities) contains a factor 
$J_{\mu\nu}(x) =\d_{\mu\nu}-2x_{\mu}x_{\nu}/x^2$ on the right hand side
for each vector index of the operator.  $J_{\mu\nu}(x)$
is the usual inversion tensor, in terms of which the Jacobian 
of the inversion is expressed
$\partial x'_{\mu}  / \partial x_{\nu} = J_{\mu \nu} (x) / x^2 $.
This prescription is essential in order to make 
vector $\Delta$-BMN operators orthonormalisable, 
see section 2 of \cite{CKTvec} for more details.

With this prescription, the two-point function \eqref{2pt} for
vector and for scalar $\Delta$-BMN operators takes the same
simple form:
\begin{equation} 
\label{2ptbar}
\langle \bar{\cO}_{\D_\alpha} (x) \cO_{\D_\beta} (0)
\rangle = 
{\d_{\alpha \beta}} \ ,
\end{equation}
which does not depend on $x$ and hence has the meaning of 
overlap of the corresponding states in the gauge theory 
Hilbert space. 

Note, however, that these $\Delta$-BMN states are the 
eigenstates of $\Delta$, 
i.e.~the eigenstates of the full interacting string Hamiltonian, 
so they cannot be identically equal to the states
from the natural string basis. The relation between the two
bases is again a linear combination
\be \label{isostdel}
\ket{s_\alpha}^{\rm SYM} \ = \ U_{\alpha \beta}\,\cO_{\D_\beta}
(x=0) \ket{0} \ ,  
\ee
with another constant matrix $U_{\alpha \beta}$.
In general, for any basis of operators 
$\tilde\cO_\alpha$ such that
\be
\label{16}
\tilde\cO_\alpha \ = \ U_{\alpha \beta}\,\cO_{\D_\beta} \ ,
\ee
the overlap is given by
\be \label{smths}
\langle \bar{\tilde\cO}_\alpha (x) \tilde\cO_\beta (0) 
\rangle \ = \ 
 U_{\beta \gamma} U^\dagger_{\gamma \alpha} \ \equiv \ 
 S_{\beta \alpha}.
 \ee
The operators $\tilde\cO_\alpha$ do not anymore have definite scaling
dimensions $\Delta$, but since they 
are expressed as a linear superposition of conformal primary operators 
which do, there
is no problem in performing the inversion required to define
$\bar{\tilde\cO}_\alpha (x)$, and the right hand side of \eqref{smths}
follows.
 
Now we describe a practical way of how to calculate 
simultaneously the overlaps and the matrix elements 
of the anomalous dimension operator 
$\delta=\Delta-\Delta_{\rm cl}$, 
where $\Delta_{\rm cl}$ is the engineering dimension. 
Let us define the barred-operator $\bar{\tilde\cO} (x)$ as the
Hermitian 
conjugation of $\tilde\cO (x)$ followed by an inversion of the 
resulting operator, defined as if it was free, i.e.~instead of
the factor $x^{2 \Delta}$ in
\eqref{scalinv} we put $x^{2 \Delta_{\rm cl}}$, such that,
\beq
\label{scalinvnew}
\bar\cO_{\Delta} (x) \, \equiv \, 
x^{2 \Delta_{\rm cl}} \, J \cdot \cO^\dagger_{\Delta}(x) \ ,
\eeq
where $J_{\mu\nu}(x)$
is the usual inversion tensor for each vector 
index (each vector impurity) of the operator.
Then the two-point function takes the form:
\be 
\label{genres}
\langle \bar{\tilde\cO}_\alpha (x) \tilde\cO_\beta (0) 
\rangle \ = \ 
 U_{\beta \gamma}\, e^{\delta_\gamma \log x^{-2}}\,
U^\dagger_{\gamma \alpha} \ = \ 
 S_{\beta\alpha} + T_{\beta\alpha} \log x^{-2} + 
\cO \, ((\log x^{-2})^2)\ .
 \ee
Here we have expanded the full result in powers of $\log x^{-2}$.
The overlap of the two states is defined as the zeroth-order term
in the expansion, 
$S_{\beta\alpha}=U_{\beta \gamma}
U^\dagger_{\gamma \alpha}$, and the matrix of anomalous
dimensions in this basis is the first order term,
\beq
T_{\beta\alpha}=U_{\beta \gamma} \delta_\gamma
U^\dagger_{\gamma \alpha} \ . 
\eeq
We note that \eqref{genres}, and hence the definitions
of the overlap and the anomalous dimension matrix, are valid to 
all orders in the gauge coupling, and so can be in principle
computed to all orders in $\l'$ and $g_2$ for any basis 
$\tilde\cO_\alpha$. 

%The matrix elements $\D_{\a \b}$ of the one-loop dilatation operator,  
%defined as
%\beq
%\D\, \cO_\a = \D_{\a \b}\cO_\b \ , 
%\eeq
%can easily be derived, at $\cO (\l')$, from 
%the expression \eqref{genres} at one loop \cite{gross2,Janik:2002bd}, 
%\beq
%\D_{\a \b} = \d_{\a \b} \,\Delta^{(0)} + (T S^{-1})_{\a \b} \ .
%\eeq

By initially relating this basis to the $\Delta$-BMN basis
we avoided all the problems of removing the `non-universal' 
$x$-dependence on the right hand side of the correlator.
Now we can forget about the $\Delta$-BMN basis and follow the
simple prescription discussed above: 
for an arbitrary basis,
the overlap matrix $S_{\beta\alpha}$ and the 
anomalous dimensions matrix $T_{\beta\alpha}$ are the zeroth
and the first term in the expansion of \eqref{genres}
in powers of $\log x^{-2}$.

We now consider the original BMN basis, for which we have
\be 
\label{origres}
\langle \bar{\cO}_\alpha (x) \, \cO_\beta  
\rangle \ = \ 
 S_{\beta\alpha} + T_{\beta\alpha} \log x^{-2} + 
 \cdots \ ,
 \ee
and relate this basis to the isomorphic to string basis 
via \eqref{isost},
\be 
\cO^{\rm string}_\beta= U_{\beta \gamma} \cO_\gamma \ \ , \qquad 
\bar{\cO}^{\rm string}_\alpha = \bar{\cO}_\delta 
U^\dagger_{\delta \alpha} \ . 
\ee
In the isomorphic to string basis (which is automatically
orthonormal, as explained earlier) we get
\be 
S^{\rm string} = \uno = U S U^\dagger \ , \qquad
T^{\rm string} =  U T U^\dagger \ .
\ee
We note that $S$ is a  Hermitian, 
positive matrix (it is a matrix of norms), 
therefore the matrix 
$S^{-1/2}$ is well-defined.%
\footnote{We would like to point our that 
this matrix $ S^{-{1\over 2}}$ appears also 
in  \cite{bits3} and \cite{SV3}.} 
$S$ is  then diagonalised by the matrix 
$U:= S^{-1/2} \cdot V$, where $V^\dagger V=\uno$:  
\beqa
S &\longrightarrow& USU^\dagger = \uno \ \ , 
\\
T &\longrightarrow& UTU^\dagger = 
V^\dagger (S^{-{1\over 2}} T S^{-{1\over 2}})V
\ \ .
\eeqa
%This transformation induces the change of basis   
%\beq
%| \a \rangle \longrightarrow| \a' \rangle =   U^{-1}_{\a' \a} | \a \rangle
%\ , 
%\eeq
%where $U^{-1} := V^{\dagger} \cdot S^{1/2}$. 
%
%g
The arbitrariness contained in $V$, which is still left at this stage 
was fixed in \cite{gross2,Gomis} 
by requiring that \eqref{ham2} holds and that the 
known three-string interaction vertex of the pp-wave 
light-cone string field theory  
\cite{SV1,SV2} is reproduced from gauge theory matrix elements 
involving BMN states (operators) with two scalar impurities. 
This condition implies $V=\uno$. 
Hence, the matrix of anomalous dimensions in the string basis
is given by
\beq
\label{Gamma}
\Gamma := T^{\rm string} = 
S^{-{1\over 2}} \, T \, S^{-{1\over 2}} \ .
\eeq
In the following sections we will show that, 
with the same choice of $V=\uno$, 
the matrix elements of $\Gamma$
between BMN operators with \\
$\bullet$ two vector impurities, \\
$\bullet$ one vector and one scalar impurity and, finally,\\ 
$\bullet$ an arbitrary number of scalar impurities,\\
precisely agree with the corresponding matrix elements 
of the interacting string Hamiltonian. 
We will consider all representations of BMN operators
with vector or scalar impurities, 
i.e.~symmetric traceless, antisymmetric and singlet.
The inclusion of vector, mixed (scalar-vector) and scalar 
BMN operators allows us to 
study the correspondence for string states in all of 
the pp-wave light-cone directions.

Other studies of the dilatation operator in gauge theory and
its interpretation in quantum mechanical models, 
which we do not pursue here,  
can be found in the recent papers
\cite{Janik,B1,SV3,B2}.

\vspace{0.5cm}
  
%%%%%%%%%%%%%%%%%%%%%%%%%%%%%%%%%%%%%%%%%%%%%%%%%%%%%%%%%%%%%%%%%%%%%%%

%%%%%%%%%%%%%%%%%%%%%%%%%%%%%%%%%%%%%%%%%%%%%%%%%%%%%%%%%%%%%%%%%%%%%
%%%%%%%%%%%%%%%%%%%%%%%%%%%%%%%%%%%%%%%%%%%%%%%%%%%%%%%%%%%%%%%%%%%%%
\section{Tests of the correspondence in the two-impurity sector:
scalar, mixed, and vector  states}
\label{2imp-sec}
In this and the following sections we will need the expressions
for the single-trace original BMN operators: 
\beqa
\label{vacuum}
\cO_{\rm vac}^J &=& {1\over \sqrt{J N_{0}^{J}}}
\Tr Z^J \ , 
\\
\label{28}
\cO_{ij , m}^J& =&\cC  
\left( \sum_{l=0}^{J}e^{2\pi i ml \over J} \Tr \left( 
\phi_i Z^l \phi_j Z^{J-l} \right)\right) \ , 
\\
\label{29}
\cO_{\mu \nu , m}^J& =&{\cC  \over 2}
\left( \sum_{l=0}^{J}e^{2\pi i ml \over J} \Tr \left[ 
( D_{\mu}Z) Z^l ( D_{\nu}Z)Z^{J-l} \right]+ 
\Tr \left[( D_{\mu}D_{\nu} Z) Z^{J+1}\right]\right) \ , 
\\
\label{pureBMNmix}
\cO_{i \mu  , m}^J &=& {\cC  \over \sqrt{2}}
\left( \sum_{l=0}^{J}e^{2\pi i ml \over J} \Tr \left[ 
\phi_i Z^l ( D_{\mu}Z)Z^{J-l} \right]+ 
\Tr \left[( D_{\mu}\phi_i ) Z^{J+1}\right]\right) \ , 
\eeqa
where  $i,j =1, \ldots, 4$,  $\m, \n =1, \ldots, 4$ label
the scalar and the vector impurities. Note that in writing
$\cO_{ij , m}^J$ and $\cO_{\mu \nu , m}^J$ 
we have taken  $i\neq j$ and $\m \neq \n$,
where the above expressions 
take the simple form \eqref{28} and \eqref{29}.
We also defined
\beq
\label{defofc}
\cC := {1\over  \sqrt{J N_0^{J+2}}}
\ , \qquad  N_0 := {g^2 \over  2}\,{N \over 4\pi^2} \ .
\eeq
The normalisation of the operators  is such that 
their two-point functions take the canonical form 
in the planar limit.
%vv
We also note that expressions for $\cO_{\mu \nu , n}^J$ and 
$\cO_{i \mu  , m}^J$ contain appropriate compensating terms
\cite{gursoy,beisert}.
These terms are required in order for the corresponding operator to 
be conformal primaries in the BMN limit.

The operators in \eqref{28}--\eqref{pureBMNmix} are the original 
BMN operators. They are related to each other 
by supersymmetry transformations \cite{beisert}. 
In order to test the correspondence, we need to use a different basis
of operators which is isomorphic to string states, as discussed earlier.
Importantly, the isomorphic to string operators
$\tilde{\cO}_{i \m , m}^J$,  $\tilde{\cO}_{\m \n  , m}^J$
are related to the  
$\tilde{\cO}_{ij , m}^J$ in exactly the same way as the original 
BMN operators are. This is because the matrix $U$ in \eqref{16} 
is a numerical matrix, i.e.~it does not contain any fields and 
does not transform under supersymmetry.
Hence, $U$ is the same for scalar, vector and mixed 
impurity BMN operators.

We will also need the expressions for the double-trace operators 
\beqa
\cT_{ij,m}^{J,y} &=& :\cO_{ij, m}^{y \cdot J }:\ 
 : O_{\rm vac}^{(1-y)\cdot J}:
 \ , 
\\ \cr
 \cT_{\m \n, m}^{J,y} &=& :\cO_{\m \n , m}^{y \cdot J }: \ 
: O_{\rm vac}^{(1-y)\cdot J}:
\ , 
\\ \cr
 \cT_{i \m ,m}^{J,y} &=& :\cO_{i \m , m}^{y \cdot J }: \ 
: O_{\rm vac}^{(1-y)\cdot J}:
\ , 
\eeqa
where $y\in (0,1)$.

{}From the three-string vertex of \cite{SV1,SV2} one extracts
the following matrix elements of the 
string Hamiltonian in the large-$\mu$ limit:
\beqa
\label{sv1}
{1\over \mu}
\langle
 \cO_{ij, m}^{J} | H_{\rm string}  | \cT_{ij, n}^{J,y}\rangle
&= & -\, C_{\rm norm} \ {\l' \over  \pi^2 y} \ \sin^2 (\pi my)
\ , 
\\ \cr
\label{sv2}
{1\over \mu}
\langle
\cO_{i \m, m}^{J} | H_{\rm string} | \cT_{i \m, n}^{J,y}\rangle
& = &  0
\ , 
\\ \cr
\label{sv3}
{1\over \mu}
\langle
\cO_{\m \n, m}^{J} | H_{\rm string} | \cT_{\m \n , n}^{J,y}\rangle
& = & C_{\rm norm}  \ {\l' \over  \pi^2 y} \ \sin^2 (\pi my)
\ , 
\eeqa
for $\m \neq \n$ and $i \neq j$.
The overall normalisation $C_{\rm norm}$ is left undetermined 
in string field theory. 
In order to get agreement with the field theory result we
will set here\footnote{$C_{\rm norm}$ is further discussed
in section 5.2, 
where we consider the case
of arbitrary many impurities, see \eqref{pre}.}  
\be
\label{38}
 C_{\rm norm}=-\frac{g_2}{2}\ \frac{\sqrt{y(1-y)}}{\sqrt{J}} \, 
  \ .
\ee
Using this normalisation, \eqref{sv1} and \eqref{sv3} become
\be
{1\over \mu}
\langle
\cO_{ij, m}^{J} | H_{\rm string}  | \cT_{ij, n}^{J,y}\rangle
=  
-{1\over \mu}
\langle
\cO_{\m \n, m}^{J} | H_{\rm string} | \cT_{\m \n , n}^{J,y}\rangle
= 
\l'\, {g_2\over \sqrt{J}} {  \sqrt{(1-y)/y}\,  \sin^2 (\pi m y) \over
2  \, \pi^2}
\ .
\ee
As mentioned earlier,
the agreement of \eqref{sv1} with the corresponding gauge theory matrix 
elements
was found in \cite{Gomis}. 
We will show that agreement with gauge theory
holds also for \eqref{sv2} and \eqref{sv3}.
%The minus sign in  \eqref{sv3} is important - it is a manifestation 
%of the $\bb{Z}_2$ non-invariance of the true pp-wave ground state. 
%We will then generalise our computations to BMN states with two scalar 
%or vector impurities in the symmetric-traceless, 
%antisymmetric and singlet representation, finding agreement 
%with string theory. 
%In this way we will exaust all the possible matrix elements 
%of two BMN operators with two impurities of any species. 
%Finally, in Section \ref{anyimp-sec} we 
%go on to construct matrix elements of the string Hamiltonian 
%involving string states with an arbitrary number of 
%scalar impurities. 
%Again, we will prove that  this matrix elements are precisely 
%matched by the corresponding field theory correlators. 

We now need the explicit form of the 
matrices $S$ and $T$ in the original BMN basis.
Both $S$ and $T$ have an expansion in powers of $g_2$, but in 
our analysis we will need their expressions only up to and including 
$\cO (g_2 )$ terms. We will also work at one loop 
in the Yang-Mills effective coupling $\l'$, where 
the matrix $T$ is of $\cO (\l' )$, whereas $S$ is of $\cO (1 )$. 
In this case, \eqref{genres} is simply
\beq
\label{stdef}
\langle \cO_{\a}(0) \bar{\cO}_\b (x)\rangle = 
%{1\over x^{2 \Delta^{(0)} }}
 S_{\a \b} + T_{\a \b} \log (x \Lambda)^{-2}
\ . 
\eeq
The pleasant fact is that 
expressions for $S$ and $T$ are closely related
and can be obtained from the coefficients
of the three-point functions, which were derived in 
\cite{BKPSS,CKTvec} for 
BMN operators with two scalar and two vector impurities, 
respectively.
We also need to know $S$ and $T$ in the case of mixed 
(i.e.~one scalar and one vector) impurities. The three-point functions 
of such BMN operators were not considered previously, 
and they will be calculated in
section \ref{sec-mix}.

The diagonal elements of $S$ and $T$ can be immediately obtained from 
\beqa
\label{andim1}
\langle
\cO_{AB, m}^J (0)\,
\bar{\cO}_{AB, n}^J (x)\rangle & = &
\d_{mn} \, \left( 1 + \l' m^2 \log (\Lambda x)^{-2}\right)
\ , 
\\ \cr
\label{andim2}
\langle
\cT_{AB, m}^{J,y} (0)\,
\bar{\cT}_{AB, n}^{J,z} (x)\rangle & = &
\d_{mn} \d_{yz} \, \left( 1 + \l' (m^2/ y^2) \log (\Lambda x)^{-2}\right)
\ . 
\eeqa
The previous expressions are valid up to $\cO (\l')$ and 
$\cO (g_2)$, and were derived originally in 
\cite{Constable1,BKPSS} for the scalar case, and in 
\cite{gursoy,gursoy2,klose} for the mixed and vector case. 

To determine the off-diagonal elements, we need to compute
the two-point correlators 
$\langle
\cT_{AB, n}^{J,y} (0)\,
\bar{\cO}_{AB, n}^J (x)
\rangle$. To this end, let us momentarily focus on the following class of 
{\it three-}point correlators, 
\beq
\label{three}
G (x_1,  x_2,  x_3 ) = 
\langle 
\cO_{AB, n}^{y\cdot J} (x_1) 
\cO_{\rm vac}^{(1-y)\cdot J}(x_2) 
 \bar{\cO}_{AB, m}^J (x_3)
\rangle \ ,  
\eeq
where $A=(i, \m )$ and $A\neq B$. 
On general grounds, 
these three-point function have the form 
\cite{Constable1,CKT,BKPSS,Constable2}
\beq \label{42}
G (x_1,  x_2,  x_3 ) =  g_2 C_{m, n y}
\left[ 1 - \l'\left( a_{m,ny} \log (x_{31} \Lambda)^2 + 
b_{m,ny} \log (x_{32} x_{31} \Lambda / x_{12}) \right) \right]
\ , 
\eeq
where $g_2 C_{m,ny}$ is the tree-level contribution, with
\beq
\label{def-C-tree}
C_{m,ny}:=
{  \sqrt{(1-y)/y}\,  \sin^2 (\pi m y) \over
\sqrt{J} \, \pi^2 (m-n / y)^2}
\ , 
\eeq
and 
the coefficients $a_{m,ny}$ and $b_{m,ny} $ must be 
calculated in perturbation theory at $\cO (\l')$.
The {\it two-}point function
$\langle \cT_{AB, n}^{J,y} (0) \, \bar{\cO}_{AB, m}^J (x) \rangle$ 
can easily be deduced from \eqref{three} by setting 
$x_{13}=x_{23} = x$ and $x_{12}= \Lambda^{-1}$ 
\cite{Constable2}, 
\beq
\label{a+b}
\langle \cT_{AB, n}^{J,y} (0)\,  \bar{\cO}_{AB, m}^J (x) \rangle
= g_2 C_{m,ny}
\left[ 1 + \l'\left( a_{m,ny} +b_{m,ny} \right)\log (x \Lambda)^{-2}  
\right]
\ .
\eeq
The matrices $S$ and $T$ are then given, up to $\cO (g_2)$, 
by
\beqa
\label{matrixS}
S &=& 
\left(\begin{tabular}{cc}
$\d_{mn}$ & $g_2\,  C_{m,qz}$  \cr \cr
$ g_2 \, C_{py,n}$ &  $\d_{pq}$
\end{tabular}\right) \ + \ \cO (g_2^2) \ = \ \uno + g_2 s +\cO (g_2^2)  \ , 
\\ \\
\label{matrixT}
T &=& \l' \left(\begin{tabular}{cc}
$m^2\, \d_{mn}$ & $g_2\,  C_{m,ny} \, ( a + b)_{m,qz}$  \cr \cr
$g_2\,  C_{py,n}\, (a+b)_{py,n}$  & $ (p^2 / y^2)\, \d_{pq} \d_{yz}$
\end{tabular}\right) \ + \ \cO (g_2^2)  
\\ \nonumber \cr
&\equiv & d\  + \ g_2 t  \ + \ \cO (g_2^2) 
\ , 
\eeqa
with 
\beqa
\label{matrix-d}
d &=& \l' \left(\begin{tabular}{cc}
$m^2\, \d_{mn}$ & $0 $  \cr \cr
$0 $  & $ (p^2 / y^2)\, \d_{pq} \d_{yz}$ 
\end{tabular}\right) \ , 
\\   
\cr
\label{matrix-t}
t &=& \l' \left(\begin{tabular}{cc}
$0 $ & $C_{m,ny} \, ( a + b)_{m,qz}$  \cr \cr
$ C_{py,n}\, (a+b)_{py,n}$  & $0$
\end{tabular}\right)   
\ .
\eeqa
It then follows that 
\beq
S^{-1/2} = \uno - g_2 (s/ 2) + \cO (g_2^2)
\eeq
diagonalises $S$ at  $\cO (g_2)$.

We now need to compute the explicit expressions for 
$a_{mn}^y$ and $b_{mn}^y$, in the scalar case, mixed (scalar-vector) case, 
and finally in the vector case.

It is easy to compute  at 
$\cO (\l')$ the coefficient $a_{mn}^y$ in planar perturbation theory,
which turns out to be
\beq
\label{a-all}
%\left[ 
a_{m,ny}  
%\right]_{\rm }
\ = \  {n^2 \over y^2} \ , 
\eeq 
independently of the type of impurity considered. 
Notice that this is exactly the $\cO (\l')$ anomalous dimension%
\footnote{It is immediate to convince oneself that the Feynman diagrams 
contributing to the $\log x_{31}^2$ part of 
$\langle 
\cO_{AB, n}^{y\cdot J} (x_1) 
\cO_{\rm vac}^{(1-y)\cdot J}(x_2) 
 \bar{\cO}_{AB, m}^J (x_3)
\rangle$ are those where the operator 
$\cO_{\rm vac}^{(1-y)\cdot J}(x_2)$ does not participate 
in the interaction,  
i.e.~they are precisely the Feynman diagrams contributing to the 
anomalous dimension of $\cO_{AB, n}^{y\cdot J} (x_1)$ - 
embedded in a three-point function.} 
of the ``small'' BMN operator at $x_1$. 

We will now explain how $ b_{m,ny} $ is determined from the coefficients
of the conformal  three-point function. First we note
that the correlator \eqref{three} does not take the conformal form
\eqref{3pt} since the original BMN operators in \eqref{42}
are not conformal primaries for $g_2 \neq 0$
due to operator mixing  \cite{Bianchi,BKPSS}. 
However,
at leading order in $g_2$, the only mixing effect which contributes to 
\eqref{3pt} 
is the presence of the double-trace corrections
in the expression for the conjugate $\Delta$-BMN operator.%
\footnote{This 
is because the double-trace corrections 
to the single-trace expression for 
an original  BMN operator is of $\cO (g_2)$, i.e.~suppressed
with $1/N$. This can be compensated by factorising the three-point function 
into a product of two two-point functions, which is possible only 
for the double-trace mixing in the operator $\bar{\cO}$.} 
Importantly, \cite{GK,CKTvec},
these mixing effects cannot affect
the remaining logarithm, $\l' \log x_{12}^2$, 
which can then be computed without taking into account mixing altogether. 
Hence, we can use the right hand side of the conformal expression
\eqref{3pt} in order to compute the coefficient $ b_{m,ny} $
in \eqref{42}.
Expanding the right-hand side of \eqref{3pt} to $\cO (\l')$, 
and equating the coefficient of the $\log x_{12}^2$ 
to the corresponding term in \eqref{three}, we  obtain 
\beq
\label{trick}
g_2 C_{m,ny} \, b_{m,ny}  = ( m^2 - n^2 / y^2) \  
 C( { \scriptstyle  A}_{n}{\scriptstyle B}_{-n},\, {\rm vac}|\, 
{\scriptstyle A}_m {\scriptstyle B}_{-m}) \ ,
\eeq
where $C( { \scriptstyle  C}_{n}{\scriptstyle D}_{-n},\, {\rm vac}|\, 
{\scriptstyle A}_m {\scriptstyle B}_{-m})$ is the coefficient 
$C_{123}$ of
the conformal three-point function 
$\langle
\cO_{C D, n}^{J_1}(x_1) 
\cO_{\rm vac}^{J_2}(x_2) {\bar{\cO}}_{A B , m}^{J  }(x_3) 
\rangle$.
We used $\Delta_1 = J_1 + 2 + \l' n^2 / y^2$, 
 $\Delta_2 = J $, and
 $\Delta_3 = J + 2 + \l' m^2 $. 

Equation \eqref{trick}
determines $b_{m,ny}$ in terms of the coefficients 
$C( { \scriptstyle  C}_{n}{\scriptstyle D}_{-n},\, {\rm vac}|\, 
{\scriptstyle A}_m {\scriptstyle B}_{-m})$
of the 
three-point function.
These coefficients for BMN operators with two scalar 
impurities, one scalar and one vector impurity,  and 
two vector impurities are given by:
\beqa 
\label{twoi1}
C( k_{n}l_{-n},\, {\rm vac}|\, i_m j_{-m})
&=&
C_{123}^{\rm vac}\frac{2\,\sin^2(\pi m y)}{y\, \pi^2 (m^2-{n^2\over y^2})^2}
\left(\delta_{i(k}\delta_{l)j}\,\,m^2+\delta_{i[k}\delta_{l]j}\,\frac{m n}{y}+
\sfrac{1}{4}\delta_{ij}\delta_{kl}\, \frac{n^2}{y^2}\right) \
\\ \cr
\label{twoimix}
C( j_{n}\n_{-n},\, {\rm vac}|\, i_m \m_{-m})
&=&
C_{123}^{\rm vac}\frac{2\,\sin^2(\pi m y)}{y\, \pi^2 (m^2-{n^2\over y^2})^2}
\delta_{ij} \delta_{\m \n}\, \, {1\over 4} \left(m + {n\over y} \right)^2 
\ , 
\\ \cr
\label{twoivec}
\nonumber
C( \r_{n}\s_{-n},\, {\rm vac}|\, \m_m \n_{-m})
&=&
C_{123}^{\rm vac}\frac{2\,\sin^2(\pi m y)}{y\, \pi^2 (m^2-{n^2\over y^2})^2}
\left(\delta_{\m(\r}\delta_{\s)\n}\,\, 
\frac{n^2}{y^2} + 
\delta_{\m[\r}\delta_{\s]\n}\,\frac{m n}{y}+
\sfrac{1}{4}\delta_{\m\n}\delta_{\r\s}m^2\, \right) \ , 
\cr \\
\eeqa
where
$C_{123}^{\rm vac}= \sqrt{JJ_1 J_2} / N= 
(g_2 /\sqrt{J})\sqrt{y(1-y)} 
$,
and the symmetric traceless and  antisymmetric traceless combinations of 
two Kronecker deltas  are defined as
\begin{equation}
\delta_{i(k}\delta_{l)j}= \sfrac{1}{2}(\delta_{ik} 
\delta_{lj} +\delta_{il} \delta_{kj}) 
- \sfrac{1}{4}\delta_{ij}\delta_{kl} \ , \quad
\delta_{i[k}\delta_{l]j}= \sfrac{1}{2}(\delta_{ik} 
\delta_{lj} -\delta_{il} \delta_{kj})  \ .
\label{sdel}
\end{equation}
The three-point function coefficient for scalars \eqref{twoi1}  
was derived  in \cite{BKPSS} (the simple case $n=0$ 
was first obtained in \cite{CKT}), 
whereas that for the vectors,  \eqref{twoivec}, 
was recently obtained  in \cite{CKTvec}.  
The three-point function coefficient 
\eqref{twoimix} for the case of mixed 
scalar-vector impurities is  a new result, and its derivation is 
presented in section \ref{sec-mix} of this paper. 

{}From 
\eqref{twoi1}--\eqref{twoivec} 
and \eqref{trick}, it is then immediate  
to derive the coefficients $ b_{m,ny}$ which correspond to 
considering scalar,  mixed, or  vector BMN operators
in \eqref{three}: 
\beqa
\label{bsc}
\left[ b_{m,ny}  \right]_{\rm scalar}&=& m^2 - {mn\over y} \ , 
\\
\label{bmix}
\left[ b_{m,ny}  \right]_{\rm scalar-vector}&=& 
{1 \over 2} \left( m^2 - {n^2 \over y^2} \right)
\ , 
\\
\label{bvec}
\left[ b_{m,ny}  \right]_{\rm vector}&=& -{n^2\over y^2}  + {mn\over y} 
\ . 
\eeqa
In conclusion, using \eqref{a+b} we get, up to $\cO (g_2)$, 
\beqa
\label{TO1}
\langle \cT_{ij, n}^{J,y} (0)\,  \bar{\cO}_{ij, m}^J (x) \rangle
&=& g_2 C_{m,ny}
\left[ 1 + \l'\left( m^2 - {mn\over y} + {n^2 \over y^2} \right)
\log (x \Lambda)^{-2}  
\right]
\ , 
\\ 
\cr 
\label{TO2}
\langle \cT_{j\n, n}^{J,y} (0)\,  \bar{\cO}_{i\m, m}^J (x) \rangle
&=& g_2 C_{m,ny}
\left[ 1 + {\l'\over 2} 
\left( m^2 + {n^2 \over y^2} \right)
\log (x \Lambda)^{-2}  
\right]\, \d_{ij} \d_{\m \n}
\ , 
\\ 
\cr 
\label{TO3}
\langle \cT_{\m \n, n}^{J,y} (0)\,  \bar{\cO}_{\m \n , m}^J (x) \rangle
&=& g_2 C_{m,ny}
\left[ 1 + \l'\left( {mn\over y} \right)\log (x \Lambda)^{-2}  
\right]
\ .
\eeqa
We will now make use of the expressions for these three correlators
to construct the matrix $T$, and therefore the matrix
$\Gamma$ dual to $H_{\rm string}^{\rm int}$, 
in the three cases of 
BMN states with (i) two scalar, (ii) one scalar  and one vector,  
and finally (iii) two vector impurities. 
These three cases are addressed separately below. 
%%%%%%%%%%%%%%%%%%%%%%%%%%%%%%%%%%%%%%%%%%%%%%%%%%%%%%%%%%%
\subsection{Matrix elements with scalar BMN states}
This case was first analysed in \cite{Gomis}, 
and we review it here for completeness. 

Substituting \eqref{a-all} and \eqref{bsc} in  \eqref{matrixT}, 
we find that the matrix $T_{\rm scalar}$  
is given, at $\cO (g_2)$, by%
\footnote{We use a somewhat simplified, but correct, notation for the 
indices of the matrices $S$ and $T$.} 
\beqa
\nonumber
T_{\rm scalar} &=& \l' 
\left(\begin{tabular}{cc}
$m^2$ & 
$g_2 C_{m,ny} \, ( m^2 - mn/y + n^2 / y^2 )$  \cr \cr
$g_2 C_{ny,m}\,( m^2 - mn/y + n^2 / y^2 ) $  & $ n^2 / y^2$
\end{tabular}\right) 
\\  \cr
&\equiv & \ 
d \ + \ g_2 \, t_{\rm scalar}
\ .
\eeqa
Multiplying it on the left and on the right by 
$S^{-1/2} = \uno - g_2 (s/ 2) + \cO (g_2^2)$ we get the expression for 
the matrix $\Gamma$ introduced in \eqref{Gamma} at $\cO ( g_2 )$:
\beqa
\Gamma_{\rm scalar} &=& 
d + g_2 \left[ t_{\rm scalar} -(1/2) \{s \, , \, d \}
\right]
\\ \nonumber \cr
&=&
\l' 
\left(\begin{tabular}{cc}
$m^2$ & 
$( g_2 C_{m,ny} /2) \, ( m - n/y )^2 $  \cr \cr
$( g_2 C_{ny,m} / 2) \, ( m - n/y )^2 $  & $ n^2 / y^2$
\end{tabular}\right) \ , 
\eeqa
from which it follows, using the definition 
\eqref{def-C-tree} of  $C_{m,ny} $,
\beq
\label{gres}
\langle \cO_{ij, m}^{J} \, | \Gamma_{\rm scalar}  | \, \cT_{ij, n}^{J,y}
\rangle = 
\l'\, {g_2\over \sqrt{J}} {  \sqrt{(1-y)/y}\,  \sin^2 (\pi m y) \over
2  \, \pi^2}
\ . 
\eeq
This result was first found in \cite{Gomis}. 
\eqref{gres} agrees with \eqref{sv1} after choosing 
the normalisation \eqref{38} for  the string result. 

%%%%%%%%%%%%%%%%%%%%%%%%%%%%%%%%%%%%%%%%%%%%%%%%%%%%%%%%%%%%%%%%%
\subsection{Matrix elements with mixed  BMN states}
In this case,  using  \eqref{a-all} and \eqref{bmix} we can determine 
the matrix $T_{\rm mixed }$ in \eqref{matrixT}
for the case of mixed impurities. It is given, at $\cO (g_2)$, by
the following expression:
\beqa
T_{\rm mixed} &=& \l' 
\left(\begin{tabular}{cc}
$m^2$ & 
$g_2 C_{m,ny} \, ( m^2 + n^2 / y^2 )/ 2$  \cr \cr
$g_2 C_{ny,m}\,( m^2 + n^2 / y^2 )/ 2  $  & $ n^2 / y^2$
\end{tabular}\right) 
\\ \nonumber \cr
&\equiv & \ 
d \ + \ g_2 \, t_{\rm mixed}
\ , 
\eeqa
where we used $a_{m,ny} + b_{m,ny}^{\rm mixed} = 
( m^2 + n^2 / y^2 )/ 2$. It then follows that 
\beq
\label{Gmixed}
\Gamma_{\rm mixed} \ = \ 
d + g_2 \left[ t_{\rm mixed} -(1/2) \{s \, , \, d \}
\right]
%\\ \nonumber \cr
%&=&
\ = \
\l' 
\left(\begin{tabular}{cc}
$m^2$ & 
$0 $  \cr \cr
$0  $  & $ n^2 / y^2$
\end{tabular}\right) \ , 
\eeq
and hence
\beq
\label{ussres}
\langle \cO_{i \m  , m}^{J} \, | \Gamma_{\rm mixed}  | \, 
\cT_{i \m , n}^{J,y} \rangle \ = \ 0 \ .  
\eeq
This verifies in gauge theory 
the vanishing of the three-string amplitude \eqref{sv2} between states with 
one scalar and one vector impurity, which was predicted in 
\cite{SV2}.
%%%%%%%%%%%%%%%%%%%%%%%%%%%%%%%%%%%%%%%%%%%%%%%%%%%%%%%%%%%%
\subsection{Matrix elements with vector BMN states}
Finally, we study the case of vector BMN impurities. 
Using  \eqref{a-all} and \eqref{bvec} we obtain  
the matrix $T_{\rm vector }$ in \eqref{matrixT}
for the case of vector impurities. At $\cO (g_2)$, it is given by:
\beqa
T_{\rm vector} &=& \l' 
\left(\begin{tabular}{cc}
$m^2$ & 
$g_2 C_{m,ny} \, ( mn/y)$  \cr \cr
$g_2 C_{ny,m}\,( mn/ y )  $  & $ n^2 / y^2$
\end{tabular}\right) 
\\ \nonumber \cr
&\equiv & \ 
d \ + \ g_2 \, t_{\rm vector}
\ , 
\eeqa
where we used $a_{m,ny} + b_{m,ny}^{\rm vector } = mn/y$. 
It then follows that 
\beqa
\label{Gvector}
\Gamma_{\rm vector} &=& 
d + g_2 \left[ t_{\rm vector} -(1/2) \{s \, , \, d \}
\right]
\\ \nonumber \cr
&=&
\l' 
\left(\begin{tabular}{cc}
$m^2$ & 
$-( g_2 C_{m,ny} /2) \, ( m - n/y )^2 $  \cr \cr
$-( g_2 C_{m,ny} /2) \, ( m - n/y )^2   $  & $ n^2 / y^2$
\end{tabular}\right) \ , 
\eeqa
from which we get
\beqa
\label{usres}
&& \langle \cO_{\m \n , m}^{J} \, | \Gamma_{\rm vector}  | \, 
\cT_{\m \n , n}^{J,y} \rangle = 
- \l'\, {g_2\over \sqrt{J}} {  \sqrt{(1-y)/y}\,  \sin^2 (\pi m y) \over
2  \, \pi^2}
\\ \nonumber  \cr
&=& - \langle \cO_{i j  , m}^{J} \, | \Gamma_{\rm scalar}  | \, 
\cT_{i j  , n}^{J,y} \rangle
\ . 
\eeqa
As advertised earlier, the  off-diagonal elements 
$\langle \cO_{\m \n , m}^{J} \, | \Gamma_{\rm vector}  | \, 
\cT_{\m \n , n}^{J,y} \rangle$
of $\Gamma_{\rm vector}$
are precisely the opposite of the corresponding elements 
$\langle \cO_{i j  , m}^{J} \, | \Gamma_{\rm scalar}  | \, 
\cT_{i j  , n}^{J,y} \rangle$
of $\Gamma_{\rm scalar}$. 
This again had been predicted in string field theory in 
\cite{SV2}. As explained in the introduction, this signals
the spontaneous breaking of $\bb{Z}_2$ symmetry in pp-wave string
theory.
%%%%%%%%%%%%%%%%%%%%%%%%%%%%%%%%%%%%%%%%%%%%%%%%%%%%%%%%%%%%
\subsection{Generalisation to all representations 
for two-impurity BMN states}
Finally, we extend our previous computations to include 
all representations of scalar and vector BMN operators 
with two impurities.

We recall here the results from the previous sections: 
 \beqa
\nonumber
&& \langle \cO_{i j , m}^{J} \, | \Gamma_{\rm scalar }  | \, 
\cT_{i j  , n}^{J,y} \rangle \ = \
-\langle \cO_{\m \n , m}^{J} \, | \Gamma_{\rm vector}  | \, 
\cT_{\m \n , n}^{J,y} \rangle \ = \  
 \l'\, {g_2\over \sqrt{J}} {  \sqrt{(1-y)/y}\,  \sin^2 (\pi m y) \over
2  \, \pi^2} \ , 
\\   \cr
&& \langle \cO_{k \m   , m}^{J} \, | \Gamma_{\rm mixed }  | \, 
\cT_{l \n   , n}^{J,y} \rangle \ = \ 0 
\ , 
\label{even}
\eeqa
($i\neq j$, $\m \neq \n$) 
which correspond to the string field theory amplitude 
 \eqref{sv1}, \eqref{sv2} and \eqref{sv3}. 
{}From these results it is immediate to obtain 
\beq
\label{samee}
\langle \cO_{\n \m , m}^{J} \, | \, \Gamma_{\rm vector}\,  |\,  
\cT_{\m \n , n}^{J,y}\rangle
= 
\langle \cO_{\m \n , m}^{J} \, | \, \Gamma_{\rm vector}\,  | \, 
\cT_{\m \n , n}^{J,y}\rangle
\ , 
\eeq
since this amounts to complex conjugate the BMN phase factor 
contained in $\cO_{\m \n , m}^{J}$, i.e.~to exchange
$m\to -m$ (same considerations apply for the scalar amplitude). 
Equation \eqref{samee} follows since the first expression in 
\eqref{even} is even in $m$. 
Therefore we can at once obtain the result for the 
symmetric-traceless and antisymmetric representations for vectors: 
\beqa
\langle \cO_{(\m \, \n ), m}^{J} \, | \, \Gamma_{\rm vector}\,  |\,  
\cT_{(\r \s), n}^{J,y}\rangle
&=& 
- \l' {g_2 \over \sqrt J}\,  {  \sqrt{(1-y)/y}\,  \sin^2 (\pi m y) \over
  \pi^2}
\delta_{\m(\r}\delta_{\s)\n} \ , 
\\ \cr
\langle \cO_{[\m \n ], m}^{J}\,  |\,  
\Gamma_{\rm vector}\,  |\,  \cT_{[\r \s ], n}^{J,y}\rangle
 &=& 0 \ , 
\eeqa
whereas for scalars, 
\beqa
\label{mse}
\langle \cO_{(i \, j ), m}^{J}\,  | \, 
\Gamma_{\rm scalar}\,  |\,  \cT_{(k l), n}^{J,y}\rangle
&=& 
 \l' {g_2 \over \sqrt J}\,  {  \sqrt{(1-y)/y}\,  \sin^2 (\pi m y) \over
  \pi^2}
\delta_{i(k}\delta_{l)j} \ , 
\\ \cr
\langle \cO_{[ij], m}^{J} \, | \, \Gamma_{\rm scalar}\,  |\,  
\cT_{[kl], n}^{J,y}\rangle
 &=& 0 \ .
\eeqa
Here we have defined 
\beq
\cO_{( \m \n )} = {1\over 2} (\cO_{ \m \n }+ \cO_{\n \m } )
- {\d_{\m \n}\over 4}   \sum_{\r} \cO_{ \r \r }
 \ , \qquad 
\cO_{[ \m \n ]} = {1\over 2} (\cO_{ \m \n }- \cO_{\n \m } ) 
 \ . 
\eeq
The vector singlet case can be treated instantly by noticing 
that the three-point function coefficient for vector singlets
is actually the same as the three-point function coefficient for 
the symmetric-traceless scalars, as it can be seen by comparing 
\eqref{twoi1} to \eqref{twoivec}.
This, together with \eqref{mse},  immediately implies that%
\footnote{The reader willing to derive explicitly   
the result \eqref{minusagain!} for  vector singlets 
%with an explicit calculation 
should 
be aware that, for singlets, \eqref{a+b} should be modified to 
$\langle \cT_{{\bf 1}, n}^{J,y} (0)\,  \bar{\cO}_{{\bf 1} , m}^J (x) \rangle
= 
g_2  C_{m,ny}
\left[ 1 + \l'\left( a_{m,ny} +b_{m,ny} \right)\log (x \Lambda)^{-2}  \right] 
+ 
g_2  C_{-m,ny}
\left[ 1 + \l'\left( a_{-m,ny} +b_{-m,ny} \right)\log (x \Lambda)^{-2}  
\right]$, and that, from the result \eqref{twoivec} derived in 
\cite{CKTvec}, 
it follows that $\left[ b_{m, ny}\right]_{{\rm vector}, \, {\bf 1}} = 
m^2 - mn / y$.
}
%%%% 
\beqa
\label{minusagain!}
&& \langle \cO_{{\rm vector}\,  {\bf 1}  , m}^{J} \, | \, 
\Gamma_{\rm vector} \,  |\,   
\cT_{{\rm vector}  \, {\bf 1}, n}^{J,y}\rangle
\ = \  
\l' {g_2 \over \sqrt J}\,  {  \sqrt{(1-y)/y}\,  \sin^2 (\pi m y) \over
  \pi^2}
\nonumber \\ \cr
&&
=\,- \langle \cO_{{\rm scalar}\,  {\bf 1}  , m}^{J} \, | \, \Gamma_{\rm scalar} 
\,  |\,   
\cT_{{\rm scalar}  \, {\bf 1}, n}^{J,y}\rangle 
\ , 
\eeqa 
where 
$\cO_{  {\bf 1}  } = ({1/ 2}) \sum_{\m}
\cO_{ \m \m   }
$. 
We notice that the result for the scalar singlet amplitude 
$\langle \cO_{{\rm scalar}\,  {\bf 1}  , m}^{J} \, | \, \Gamma_{\rm scalar} 
\,  |\,   
\cT_{{\rm scalar}  \, {\bf 1}, n}^{J,y}\rangle$
agrees with the result  found in \cite{Gomis2}.
The opposite sign in \eqref{minusagain!} for the vector singlet compared
to the scalar singlet case is again a manifestation 
of the (spontaneously broken)  $\bb{Z}_2$ symmetry in 
pp-wave string theory.
%Equation \eqref{minusagain!} is one of our principal results. 

%%%%%%%%%%%%%%%%%%%%%%%%%%%%%%%%%%%%%%%%%%%%%%%%%%%%%%%%%%%%
%%%%%%%%%%%%%%%%%%%%%%%%%%%%%%%%%%%%%%%%%%%%%%%%%%%%%%%%%%%%
\section{A technical aside: three-point function with mixed impurities} 
\label{sec-mix}
In this section we derive the expression \eqref{twoimix}.
The reader not interested in the actual calculation 
can turn a few pages and proceed to the next section.

Here we would like to compute 
the coefficient of the three-point function of one vacuum
operator \eqref{vacuum}  and 
two conformal primary $\Delta$-BMN operators with 
one scalar and one vector impurity, 
\beq
\label{dots}
\tilde{\cO}_{i \mu  , m}^J = \cO_{i \mu  , m}^J + \cdots \ , 
\eeq
where $\cO_{i \mu  , m}^J $ is defined in \eqref{pureBMNmix}.
The operator $\tilde{\cO}_{i \mu  , m}^J$ has a definite scaling dimension,
$\Delta_{n} = \Delta_{\rm cl} + \delta_n$, 
which implies that the single-trace expression 
$\cO_{i \mu  , m}^J$ on the right hand side of
\eqref{dots} must be accompanied by  multi-trace corrections 
(and other mixing effects) at higher orders in $g_2$
\cite{Bianchi,BKPSS}.
The dots on the right hand side  of \eqref{dots} 
stand for these corrections. 

Nevertheless, 
our strategy is to study  the three-point correlator
of the {\it original} BMN operators $\cO_{i \mu  , m}^J $, 
\beqa
\label{corr}
&& \langle 
\cO_{j \n, n}^{y\cdot J} (x_1) 
\cO_{\rm vac}^{(1-y)\cdot J}(x_2) 
 \bar{\cO}_{i \m, m}^J (x_3)
\rangle  = 
\\ \nonumber \cr
&& g_2 C_{m, n y}
\left[ 1 - \l'\left( a_{m,ny} \log (x_{31} \Lambda)^2 + 
b_{m,ny}^{\rm mixed} 
\log \left|{x_{32} x_{31} \Lambda \over x_{12}}\right| 
\right) \right]\, \delta_{\m \n} \delta_{i j }
\ , 
\eeqa
and to focus on the computation of the coefficient $b_{m,ny}^{\rm mixed} $ 
of the 
$\log x_{12}$.
This is because, for reasons explained in the paragraph
below \eqref{a-all},
this coefficient 
can be computed without taking into account mixing altogether, 
and is directly related to the coefficient of the three-point function 
of conformal primary BMN operators with mixed impurities  through
\eqref{trick}. Therefore, from now on we will work with 
the original BMN operator \eqref{pureBMNmix}.

The mixed BMN operator in 
\eqref{pureBMNmix} contains two terms: a ``pure'' BMN part
and a compensating term, first and second term on the right hand side of 
\eqref{pureBMNmix}, respectively. 
The Feynman diagrams contributing to  
the Green's function in \eqref{corr} can be divided into 
two classes: those obtained by taking only the pure BMN part of 
$\cO_{j \n, n}^{y\cdot J} (x_1) $ and $\bar{\cO}_{\ i \m, m}^J (x_3)$
and those where the compensating part is taken (in one or both operators). 
As already explained, we will focus only on diagrams which can produce 
a $\log x_{12}$ dependence, and 
both classes of diagrams contribute to the coefficient 
$b_{m,ny}^{\rm mixed} $ in \eqref{corr}. For the sake of clarity, 
we quote here the  results from these two classes of diagrams: 
\beqa
\left[ b_{m,ny}^{\rm mixed} \right]_{\rm  BMN} &=&
m\left(m-{n\over y}\right) \ , 
\\
\left[ b_{m,ny}^{\rm mixed} \right]_{\rm comp} &=&
-{1\over2} \left(m-{n\over y}\right)^2 \ .  
%\\
%\left[ b_{m,ny}^{\rm mixed} \right]_{\rm total} &=&
%\left[ b_{m,ny}^{\rm mixed} \right]_{\rm BMN} +
%\left[ b_{m,ny}^{\rm mixed} \right]_{\rm comp} = 
%{1\over2}\left(m^2-{n^2\over y^2}\right) \ , 
\eeqa
The total result 
\beq
\left[ b_{m,ny}  \right]_{\rm scalar-vector}
=\left[ b_{m,ny}^{\rm mixed} \right]_{\rm  BMN}+
\left[ b_{m,ny}^{\rm mixed} \right]_{\rm comp} \ = \
{1\over2}\left(m^2-{n^2\over y^2}\right) \ , 
\eeq
 was anticipated in \eqref{bmix}.
We now compute  separately these two classes of diagrams.
Our notation and conventions are summarised in Appendix B.
%%%%%%%%%%%%%%%%%%%%%%%%%%%%%%%%%%%%%%%%%%%%%%%%%%%%%%%%%%%%
\subsection{Diagrams originating from the ``pure'' BMN parts}
We consider first the diagrams where 
the scalar impurity interacts.
These are represented in Figure 1. 
\begin{figure} [ht]
\label{fig1}
\psfrag {fi} {\LARGE ${\phi_i}$}
\psfrag {zb} {\LARGE${\bar{Z}}$}
\psfrag {dmzb} {\LARGE${\scriptstyle\overline{\partial_{\mu}Z}}$}
\psfrag {x1}{\LARGE$x_1$}
\psfrag {x2}{\LARGE$x_2$}
\psfrag {z}{\LARGE$Z$}
\psfrag {1a}{{\LARGE 1a}}
\psfrag {1b}{{\LARGE 1b}}
\psfrag {1c}{{\LARGE 1c}}
\psfrag {1d}{{\LARGE 1d}}
\psfrag {1e}{{\LARGE 1e}}
\psfrag {1f}{{\LARGE 1f}}
\begin{center}
{\scalebox{0.5}{
\includegraphics{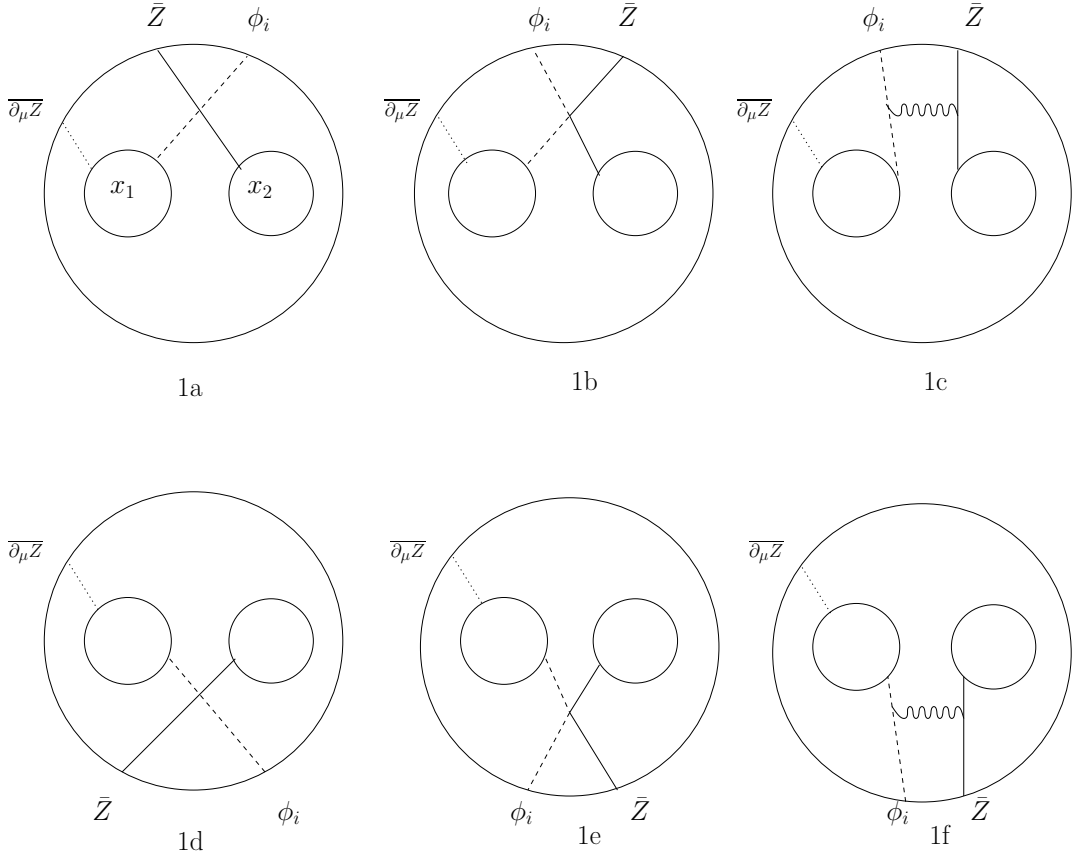}}
}
\end{center}
\caption{Diagrams with scalar impurity interacting.
Diagrams 1a and 1d have positive signs, all the others 
 have negative signs.
}
\end{figure}
The result for these diagrams is: 
\beq
\label{primo}
\left({2 \over g^2}\right) 
\left({g^2 \over 2}\right)^4 \cdot 2(P_I - P_{II} + \bar{P}_{I} 
- \bar{P}_{II} )\cdot (2 \delta_{\m \n}) \delta_{i j } \cdot X
\ .
\eeq
The first term on the right hand side of \eqref{primo} comes from the
diagram 1a (the coefficient of 2 is easily seen from 
$-V_F$ in \eqref{effe}), 
the second term is  the sum of diagrams  1b and 1c. 
The relative sign is also immediately  seen from the commutators in 
$-V_F$. 
We have taken into account the fact that 
diagrams 1b and 1c give the same contribution.%
\footnote{This is a simple corollary of the cancellation of
D-terms  against gluon interactions and self-energies in 
three-point functions of BMN operators at $\cO (\l' )$ 
(in the complex basis) \cite{D'Hoker:1998tz,Constable1,CKT,BKPSS}. 
In our  case, self-interactions diagrams do not participate since 
they cannot generate $\log x_{12}^2$ terms at order $\cO (\l' )$.}

The third and fourth term in \eqref{primo} come 
from the mirror diagrams  1d, 1e and 1f, where the $\phi$
interaction is now at the bottom. 
The factor $2 \delta_{\m \n}$ comes from the free contraction%
\footnote{\label{footnote}For an extensive discussion 
of the treatment of BMN 
operators with vector impurities, the reader is referred to 
\cite{CKTvec}. Free contractions of vector impurities are discussed 
in Eq.~(34) of that paper, and the main results can be summarised as
$\langle \overline{ D_\m Z} \, \, D_\n Z \rangle_{\rm free} \, = \,  
2 \, \d_{\m \n}, $
and $\langle \overline{ Z} \, \, D_\n Z \rangle_{\rm free} \, = \,  0 $.
}
of the $\overline{D_{\m} Z}$ impurity   with the  $D_{\n}Z$ impurity.
The coefficients $P_{I}$ and $P_{II}$ come from summing over 
the BMN phase factors, and their expressions are summarised in Appendix C.
Mirror diagrams are associated with the complex conjugate coefficients
 $\bar{P}_{I}$ and $\bar{P}_{II}$. 
Finally, the function $X$ is defined in \eqref{X} of Appendix D.

There are additional diagrams, drawn in Figures 2 and 3, where 
the interaction involves now the vector impurity. 
\begin{figure}[ht]
\psfrag{phi}{\Large $\overline{\partial_\m Z}$}
\psfrag{psi}{\Large $\phi_i$}
\psfrag{Zb}{\Large $\bar{Z}$}
\psfrag{k}{\Large $k$}
\psfrag{l}{\Large $l$}
\psfrag{x1}{\Large $x_1$}
\psfrag{x2}{\Large $x_2$}
\psfrag{(1)}{{\LARGE 2a}}
\psfrag{(2)}{{\LARGE 2b}}
\psfrag{(3)}{{\LARGE 2c}}
\begin{center}
{\scalebox{0.5}{
\includegraphics{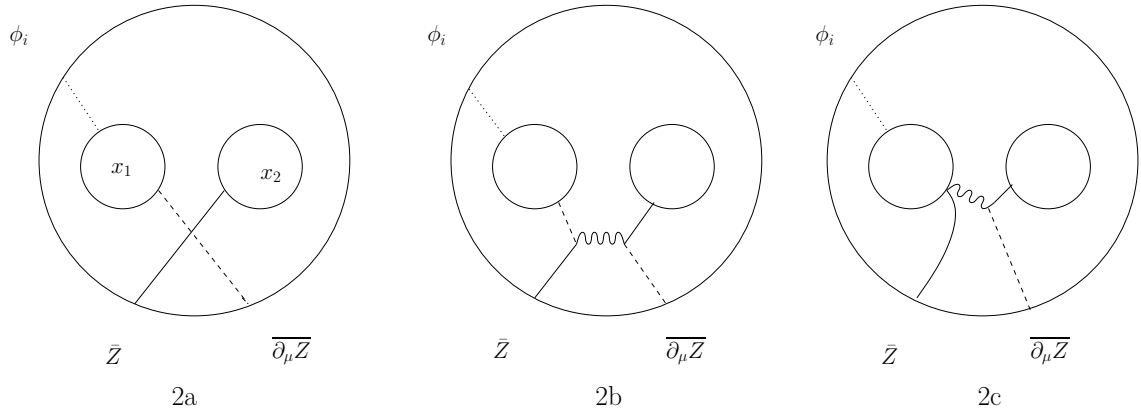}}
}
\end{center}
\caption{Diagrams with vector impurity interacting  associated to $P_{I}$.}
\label{fig2}
\end{figure}

\begin{figure}[ht]
\psfrag{phi}{\Large$\overline{D_\m Z}$}
\psfrag{psi}{\Large$\overline{D_\n Z}$}
\psfrag{k}{\Large$k$}
\psfrag{l}{\Large$l$}
\psfrag{x1}{\LARGE$x_1$}
\psfrag{x2}{\LARGE$x_2$}
\psfrag{Z}{\Large$Z$}
\psfrag{Zb}{\Large $\bar{Z}$}
\psfrag{Zc}{\Large $\overline{\partial_{\m}Z}$}
\psfrag{dZ}{\Large $\partial_{\n} Z$}
\psfrag{DZ}{\Large $D_{\n} Z$}
\psfrag{a}{\Large $\phi_i$}
\psfrag{(1)}{{\Large 3a}}
\psfrag{(2)}{{\Large 3b}}
\psfrag{(3)}{{\Large 3c}}
\psfrag{(4)}{{\Large 3d}}
\psfrag{(5)}{{\Large 3e}}
\psfrag{(6)}{{\Large 3f}}
\begin{center}
{\scalebox{0.60}{
\includegraphics{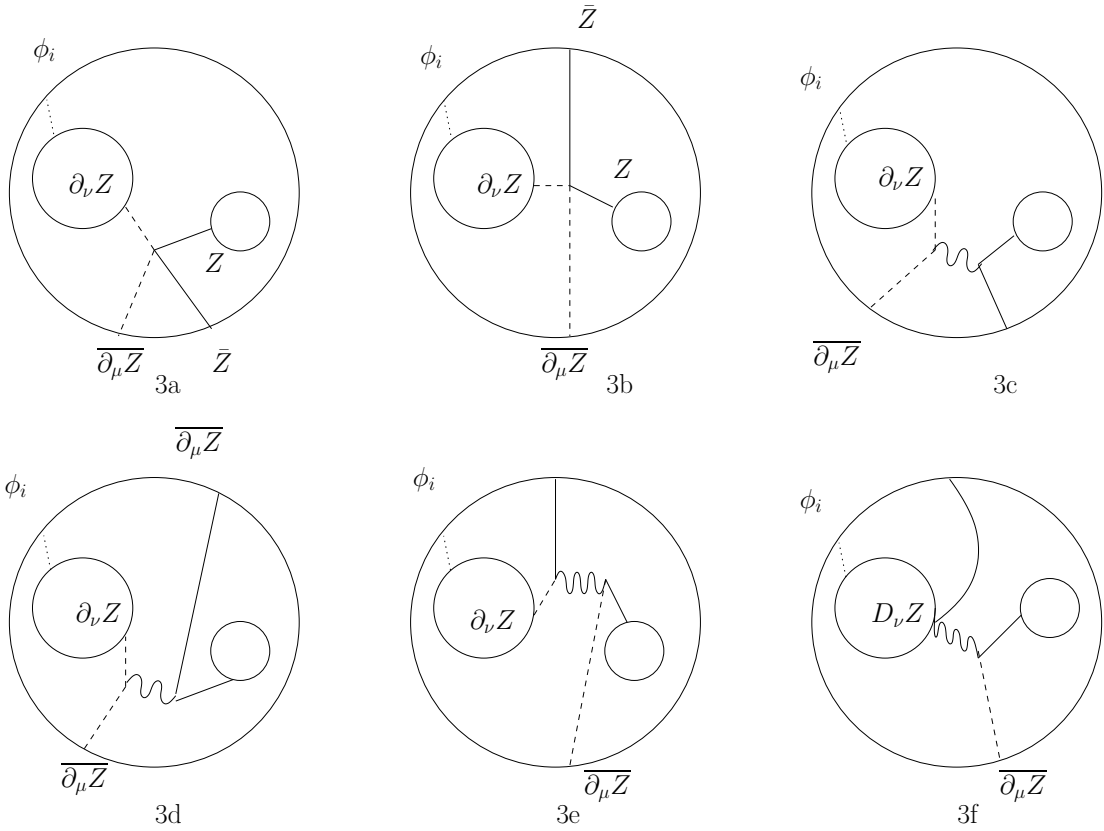}}
}
\end{center}
\caption{Diagrams with vector impurity interacting  
associated to $P_{II}$.}
\label{fig3}
\end{figure}
These  diagrams are identical to those in Figure 5 and 6 
of \cite{CKTvec}, with the only modification that 
the non-interacting impurity 
is now a scalar impurity (whereas in Figure 5 and 6 of \cite{CKTvec} 
it was a vector impurity). 
We will not compute again these diagrams, and instead borrow 
the result from  \cite{CKTvec}. Their contribution 
 turns out to be precisely the same of the contribution  \eqref{primo}
from the diagrams where the scalar impurity interacts.

The  final result for the pure BMN diagrams is therefore: 
\beq
\label{secundo}
\left({2 \over g^2}\right) 
\left({g^2 \over 2}\right)^4 \cdot 8(P_I - P_{II} + \bar{P}_{I} 
- \bar{P}_{II} )\cdot \delta_{\m \n} \delta_{i j } \cdot X
\ .
\eeq
This quantity has still to be multiplied by 
the normalisations of the operators, in which we include 
an extra factor of 
$J_2=(1-y)\cdot J$ coming from inequivalent Wick contractions 
of $\cO_{\rm vac}$ with the rest, 
\beq
\label{normop3}
{1 \over \sqrt{J}} {\sqrt{1-y}\over \sqrt{y}} 
\left({1\over \sqrt{2}}\right)^2 
%\left({g^2 \over 2}\right)^{-2}  
\ .
\eeq

%%%%%%%%%%%%%%%%%%%%%%%%%%%%%%%%%%%%%%%%%%%%%%%%%%%%%%%%%%%%
\subsection{Diagrams from compensating terms}
The compensating term is present in both operators at $x_3$ and $x_1$, 
therefore there are three subclasses of diagrams: 
(i) diagrams with compensating term  in the ``external''  operator 
at $x_3$ and 
pure BMN part in the ``internal'' (small)  operator at $x_1$; 
(ii) diagrams where the compensating term 
of both operators at $x_1$ and $x_3$ is considered; and
(iii) diagrams where the compensating term of 
the small operator at $x_1$  and the pure BMN part of 
the operator at $x_3$ are taken.

Each diagram in class (i) vanishes separately, since the only way 
to contract the impurity $D_\n Z$ in $\cO_{j \n, n}^{y\cdot J} (x_1) $ is 
with a $\bar{Z}$ in $\bar{\cO}_{i \m, m}^{J} (x_3) $, 
and this contraction vanishes (see footnote \ref{footnote}).
Moreover , it is not difficult to see that 
the total contribution of the diagrams in class (ii) 
vanishes. Hence we are left with diagrams in class (iii), which 
we now discuss.

We start by considering diagrams without gluons. 
\begin{figure} [ht]
\label{fig4}
\psfrag{fi}{\Large ${\phi_i}$}
\psfrag{zb}{\Large ${\bar{Z}}$}
\psfrag{k}{\Large ${%\scriptstyle
\overline{\partial_{\mu}Z}}$}
\psfrag{dfi}{\Large ${%\scriptstyle
{\partial_{\nu}\phi_j}}$}
\psfrag{z}{\Large $Z$}
\psfrag{4a}{{\LARGE 4a}}
\psfrag{4b}{{\LARGE 4b}}
\psfrag{4c}{{\LARGE 4c}}
\begin{center}
{\scalebox{0.5}{
\includegraphics{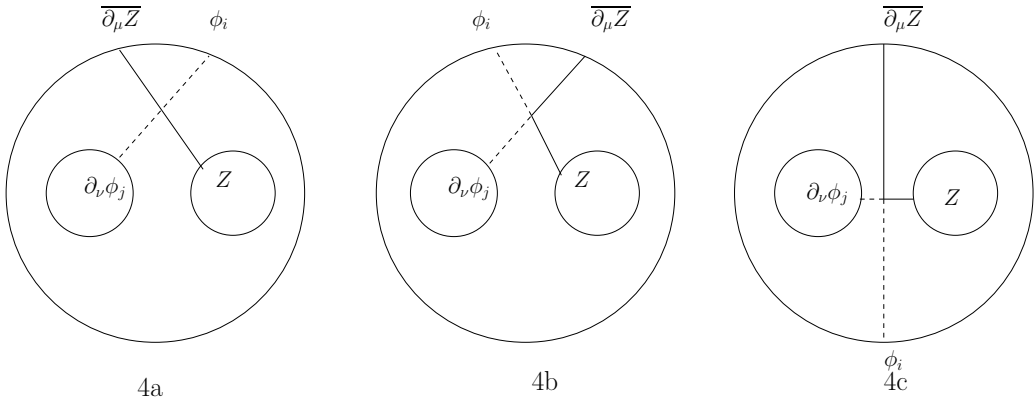}}
}
\end{center}
\caption{ Diagrams originating from the compensating term in the 
`internal' operator (at $x_1$). In these diagrams no gluons 
are emitted or exchanged.}
\end{figure}
The first three diagram are represented in Figure 4, 
and their contribution is 
\beq
\label{ter}
\left({2 \over g^2}\right) 
\left({g^2 \over 2}\right)^4 \cdot 
\left( 2X - X - X \cdot \bar{q}^{J_2} \right) \,
\delta_{\m \n} \delta_{i j } 
\ , 
\eeq
where  
$q= \exp (2\pi m / J)$ is the phase factor of the BMN operator at 
$x_3$.
The first term in the right hand side of \eqref{ter} comes from 
the first diagram in Figure 4.  This diagram is equal to the first diagram 
in Figure 5 of \cite{CKTvec}, from which we borrowed the result. 
The factor of 2 is easily seen from the term 
$\Tr ( 2\, Z \phi_i \bar{Z} \phi_i )$
in $-V_{F}$, see \eqref{effe}. The opposite sign of the second term 
in \eqref{ter} is also seen from the term 
$ -\Tr (\phi_i \phi_i  Z \bar{Z} ) $ in $-V_{F}$. The third term comes 
from the term $ -\Tr (\phi_i \phi_i   \bar{Z} Z ) $ in $-V_{F}$, and 
carries a BMN phase factor $\bar{q}^{J_2}$. 
There are also mirror diagrams, where the interaction occurs at the bottom 
of the diagram (similarly to the fourth, fifth and sixth diagram in Figure 1). 
As usual, their effect is to add the complex conjugate of the previous result, 
so that the final result for diagrams without gluons is: 
\beq
\label{ter2}
\left({2 \over g^2}\right) 
\left({g^2 \over 2}\right)^4 \cdot 
2X \cdot \left( 1-\cos \, (2\pi m y)  \right) 
\, \delta_{\m \n} \delta_{i j } 
\ . 
\eeq
We now consider diagrams where a gluon is emitted from 
the covariant derivative ${D_{\n} \phi_j}$  
at $x_1$. 
These gluon emission diagrams are represented in Figure 5. 
The total result for them is: 
\beq
\label{quater}
\left({2 \over g^2}\right) 
\left({g^2 \over 2}\right)^4 \cdot 
\left( 3 X - 3 X\cdot \bar{q}^{J_1} \right) \,
\delta_{\m \n} \delta_{i j } 
\ . 
\eeq
The first term on the right hand side of \eqref{quater} 
corresponds to the first diagram in Figure 3. 
\begin{figure} [ht]
\label{fig5}
\psfrag{fi}{\Huge ${\phi_i}$}
\psfrag{zb}{\Huge${\bar{Z}}$}
\psfrag{dzb}{\Huge${\scriptstyle\overline{\partial_{\mu}Z}}$}
\psfrag{dfi}{\Huge${\scriptstyle{D_{\nu}\phi_j}}$}
\psfrag{z}{\Huge$Z$}
\psfrag{5a}{{ \mbox{\Huge 5a}}}
\psfrag{5b}{{\Huge 5b}}
\begin{center}
{\scalebox{0.45}{
\includegraphics{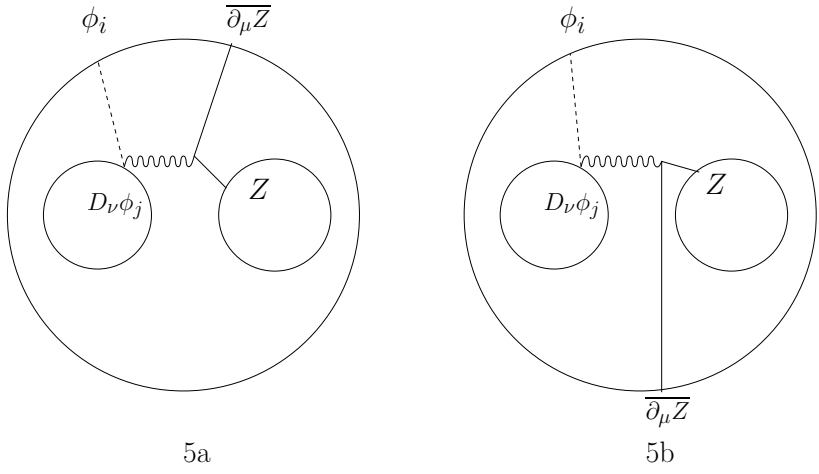}}
}
\end{center}
\caption{Gluon emission diagrams originating 
from the compensating term in the internal operator.}
\end{figure}
This diagram was computed in \cite{CKTvec} (it is the third diagram 
in Figure 5), from which we took the result. The only difference is that 
in the present case it is accompanied by phase factor equal to 1.
The second term come from the second diagram in Figure 3, and carries a 
BMN phase factor equal to 
$\bar{q}^{J_1}$. 
Again, there are also mirror diagrams, 
where the interaction occurs at the bottom 
of the diagram.
Their effect is to add the complex conjugate of the previous result, 
so that the final result for diagrams with gluon emission is: 
\beq
\label{quater2}
\left({2 \over g^2}\right) 
\left({g^2 \over 2}\right)^4 \cdot 
6X \cdot \left( 1-\cos \, (2\pi m y)  \right) 
\, \delta_{\m \n} \delta_{i j } 
\ . 
\eeq
Finally, we have to consider gluon interaction diagrams. 
These are depicted in Figure 6 and, as before, there are 
also mirror diagrams, where the interaction 
occurs at the bottom of each diagram. 
\begin{figure} [ht]
\label{fig6}
\psfrag {fi} {\Huge${\phi_i}$}
\psfrag {zb} {\Huge${\bar{Z}}$}
\psfrag {dzb} {\Huge${\scriptstyle\overline{\partial_{\mu}Z}}$}
\psfrag {dfi} {\Huge${\scriptstyle{\partial_{\nu}\phi_j}}$}
\psfrag {z}{\Huge$Z$}
\psfrag {6a}{{\Huge 6a}}
\psfrag {6b}{{\Huge 6b}}
\begin{center}
{\scalebox{0.4}{
\includegraphics{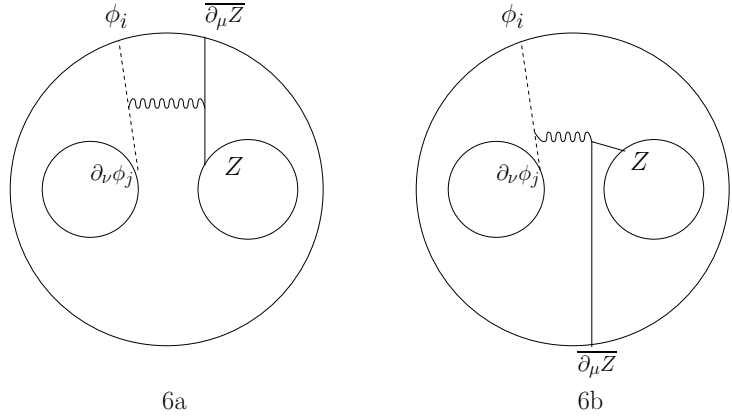}}
}
\end{center}
\caption{Gluon interaction diagrams, 
from the compensating term in the internal operator.}
\end{figure}
However, each of the diagrams vanishes separately
(this is to be contrasted with the case of scalar interactions, 
where gluon interactions double up the result for the interaction
from the scalar potential, 
as discussed in the previous subsection). 
This was shown again in \cite{CKTvec} 
(second diagram in Figure 5 of that paper). 

Adding \eqref{ter2} and \eqref{quater2}  we get the total contribution
of the compensating term diagrams,  
\beq
\label{quater3}
\left({2 \over g^2}\right) 
\left({g^2 \over 2}\right)^4 \cdot 
16X \cdot \sin^2 \, (\pi m y)  
\, \delta_{\m \n} \delta_{i j } 
\ . 
\eeq
The result \eqref{quater3}  has still to be multiplied by 
the normalisations of the operators \eqref{normop3}.
%%%%%%%%%%%%%%%%%%%%%%%%%%%%%%%%%%%%%%%%%%%%%%%%%%%%%%%%%%%
\subsection{Summary: the result for mixed impurities}
We add the results \eqref{secundo}
and \eqref{quater3}, and use 
\eqref{totalpf} of Appendix C, to get the total result
\beq
\left({g^2 \over 2}\right)^3 \cdot (-16\, X) 
 \cdot ( \delta_{\m \n} \delta_{i j } ) \,  \cdot\, 
{ m +  n/  y  \over {m - n/  y}}\,  \sin^2 \pi m y
\ . 
\eeq
Multiplying this result by the normalisation \eqref{normop3}, and
amputating a factor of $\left[ ( g^2 / 2)\,  \D (x_{13})\right]^2$ 
we obtain 
\beqa
\nonumber
&& 
\left. 
\langle 
\cO_{j \n, n}^{y\cdot J} (x_1) 
\cO_{\rm vac}^{(1-y)\cdot J}(x_2) 
 \bar{\cO}_{i \m, m}^J (x_3)
\rangle 
\right|_{\log x_{12} \ {\rm term}} = 
\\ \nonumber \cr
&& {1 \over \sqrt{J}} {\sqrt{1-y}\over \sqrt{y}} 
\left({1\over \sqrt{2}}\right)^2 \left({g^2 \over 2}\right)\cdot
\left( -16 \, { m +  n/  y  \over {m - n/  y}}\,  \sin^2 \pi m y
\right) {\log (x_{12} \L )^{-1}\over 8 \pi^2} 
\delta_{\m \n} \delta_{i j }\ = \ 
\\  \cr
&& - \l'  g_2 \, C_{m, ny} \cdot 
{1\over 2} \left( m^2 - {n^2 \over y^2} \right) 
\cdot \log (x_{12} \L )^{-1}\delta_{\m \n} \delta_{i j }
\ , 
\label{ah}
\eeqa
where $C_{m, ny}$ is defined in   \eqref{def-C-tree}. 
Equation \eqref{ah}  is the principal result of this section. 

The coefficient 
$\left[ b_{m,ny}  \right]_{\rm scalar-vector}$ 
in \eqref{bmix}  immediately follows by 
comparing \eqref{ah} to \eqref{corr}.  
Finally, the three-point function coefficient 
\eqref{twoimix} for mixed impurities is obtained  
from \eqref{bmix} and \eqref{trick}.

%%%%%%%%%%%%%%%%%%%%%%%%%%%%%%%%%%%%%%%%%%%%%%%%%%%%%%%%%%%%

%%%%%%%%%%%%%%%%%%%%%%%%%%%%%%%%%%%%%%%%%%%%%%%%%%%%%%%%%%%%%%%%%%%%%%%%%%
\section{The correspondence for an 
arbitrary number of scalar impurities}
\label{anyimp-sec}
In this section we shall evaluate the coefficients of three-point 
functions of $\Delta$-BMN operators with arbitrary number of 
scalar impurities, and use this information to derive the 
single- double-trace two-point function of operators
with an arbitrary number of scalar impurities.

\subsection{The results in field theory}
Every  BMN operator with an arbitrary number of impurities can
be decomposed into two pieces. The pure BMN part, which
contains no $\bar{Z}$, and the compensating part, which 
contains $\bar{Z}$. In order to make the 
structure of the general BMN operator clear, 
let us consider the example of an
operator with three impurities. 
The pure BMN part  consists of two terms:
\bea 
\nonumber
\cO_{\rm pure}=\sum_{0\leq l_{2}\leq l_{3}}q_{2}^{l_{2}}q_{3}^{l_{3}}
 \, \Tr (\phi_{1}Z^{l_{2}}\phi_{2}Z^{l_{3}-l_{2}}\phi_{3}Z^{J-l_{3}})
+\sum_{0\leq l_{3}\leq l_{2}}q_{2}^{l_{2}}q_{3}^{l_{3}}
 \, \Tr(\phi_{1}Z^{l_{3}}\phi_{3}Z^{l_{2}-l_{3}}\phi_{2}Z^{J-l_{2}})
\ .
\\ 
\label{95}
\eea
In all cases  $\phi_1$ is positioned first in the trace, 
while we have to sum over all the different orderings of the 
remaining  
impurities. Let us denote by  $\phi_{p(i)}$ the impurity
which sits in the $i^{\rm th}$ position of a specific ordering
of the impurities, and $l_{p(i)}$ the number of $Z$ fields 
between 
$\phi_1$ and $\phi_{p(i)}$ (in the example given above
$p(2)=2$, $p(3)=3$ for the first trace 
while  $p(2)=3$, $p(3)=2$ for the second trace). 
Next we consider the compensating
terms. In our example, these should be written as
\bea 
\label{96}
\cO_{\rm comp}=-\d _{\phi_{2}\equiv \phi_{3}}\sum_{0\leq l_{2}}
 (q_{2}q_{3})^{l_{2}}\, \Tr(\phi_1 Z^{l_{2}} \bar{Z}Z^{J-l_{2}})
-\d _{\phi_{1}\equiv \phi_{2}}\sum_{0\leq l_{3}}
 q_{3}^{l_{3}}\, \Tr(\bar{Z}Z^{l_{3}}\phi_3 Z^{J-l_{3}})\nonumber\\
-\d _{\phi_{1}\equiv \phi_{3}}\sum_{0\leq l_{2}}
 q_{2}^{l_{2}} \, \Tr(\bar{Z} Z^{l_{2}}\phi_2 Z^{J-l_{2}})
\ . 
\eea
In other words, whenever two impurities in  $\cO_{\rm pure}$ are
of the same flavour, we add a compensating term where these two 
impurities are replaced by $\bar{Z}$.

With this example in mind, it is not difficult to 
write down the most  general form of an operator with $n$ impurities, 
\bea
\cO_{\{n_{i}\}} \equiv \frac{1}{\sqrt{J^{n-1} N^{J+n}}}
\sum_{p= {\rm perm}\{2, \ldots ,n\}} \cO_{\{n_{i}\}}^{1p} 
\ , 
\eea
where $i=1, \ldots ,n$ and
\bea
\label{pure}
\cO_{\{n_{i}\}\rm pure}^{1p(2) \ldots p(n)}=
\sum_{0\leq l_{p(2)}\leq l_{p(3)} \ldots \leq l_{p(n)}}
\prod_{i=2}^{n}q_{p(i)}^{l_{p(i)}} \, 
\Tr(\phi_{1}Z^{l_{p(2)}}\phi_{p(2)}
Z^{l_{p(3)}-l_{p(2)}}\phi_{p(3)} \ldots \phi_{p(n)}Z^{J-l_{p(n)}})
\ , 
\nonumber\\
\eea
\bea 
\label{comp}
\cO_{\{n_{i}\}\rm comp}^{1p(2) \ldots p(n)}=
- \frac{1}{2}\sum_{k=2}^{n-1} \d _{\phi_{p(k)}\equiv \phi_{p(k+1)}}
\sum_{0\leq l_{p(2)} \ldots \leq l_{p(k)}\leq l_{p(k+2)}\ldots \leq l_{p(n)}}
\prod_{i=2,i\neq p(k+1)}^{n}q_{p(i)}^{l_{p(i)}}q_{p(k+1)}^{l_{p(k)}}
\nonumber\\
\, \Tr(\phi_{1}Z^{l_{p(2)}}\phi_{p(2)}...Z^{l_{p(k)}-l_{p(k-1)}}
\bar{Z}Z^{l_{p(k+2)}-l_{p(k)}}\phi_{p(k+2)}\ldots\phi_{p(n)}Z^{J-l_{p(n)}})
\nonumber\\
-\d _{\phi_{1}\equiv \phi_{p(2)}}\sum_{0\leq l_{p(3)}\ldots\leq l_{p(n)}}
\prod_{i=3}^{n}q_{p(i)}^{l_{p(i)}}
\, \Tr(\bar{Z}Z^{l_{p(3)}}\phi_{p(3)} \cdots
\phi_{p(n)}Z^{J-l_{p(n)}})
\ . 
\eea
The origin of the $\frac{1}{2}$ in front of the first term 
of $\cO_{\{n_{i}\}{\rm comp}}^{1p(2)\ldots p(n)}$ is quite clear. 
It comes from the fact
that we have counted twice the same term, 
since we have two orderings where
$\phi_{p(i)}$ and  $\phi_{p(i+1)}$ coincide. The one with $\phi_{p(i)}$
coming first and $\phi_{p(i+1)}$ following and vice versa.
Finally, we note that in principle more compensating 
terms should be added to the right hand side of \eqref{comp} when 
two or more pairs of impurities coincide. These terms are 
irrelevant for our purposes in the BMN limit.

In the above expression,  $\phi_{p(i)} \in \{ \phi_{1}, 
\phi_{2},\phi_{3},\phi_{4}\}$. This makes the meaning of the 
Kronecker $\d $-symbols functions obvious.  
It should also be noted that the operator given above is normalised 
so that its 
two point function is one at the free theory level.%
\footnote{Strictly speaking this is true 
only when all the $q$'s which correspond to a particular $\phi$, 
say $\phi_{1}$, are different.
 This is the case that we are going to consider. 
However what follows can be applied with slight 
modifications to the case where two or more of the $q$'s are the same.} 

As in section 3, we will need the expressions for double-trace operators,
\be \label{tnewdef}
\cT_{{\{n_{i}\}} {\{k_{i}\}}}^{J,y} =\, :\cO_{\{n_{i}\}}^{J\cdot y}: \,
:\cO_{\{k_{i}\}}^{J\cdot (1-y)}: \ .
\ee
On general grounds, 
the two-point function of the double- and single-trace BMN operators
takes the form
\beq
\label{a+b+c}
\langle \cT_{{\{n_{i}\}} {\{k_{i}\}}}^{J,y} (0)\,  
\bar{\cO}_{\{m_{i}\}}^J (x) \rangle
= g_2 C_{\rm free}
\left[ 1 + \l'\left( a +b +c \right)\log (x \Lambda)^{-2}  
\right]
\ .
\eeq 
In \eqref{a+b+c} we have suppressed the indices of $a$, $b$
and $c$. 
Here  $ \bar{\cO}_{\{m_{i}\}}^J$ contains $p_3$ impurities, 
whereas the two single-trace expressions in 
$\cT_{{\{n_{i}\}} {\{k_{i}\}}}^{J,y}$
contain $p_1$ and $p_2$ impurities,  respectively.

Compared to \eqref{a+b}, the above equation  contains 
a  new coefficient, $c$.
This is due to the fact that the second operator 
on the right hand side of \eqref{tnewdef} is no 
longer just the vacuum, but instead is 
a generic string state. This results 
in an additional logarithmic part for the three-point 
function \eqref{42}, i.e.~$c \cdot \log(x_{32}\L)^2)$.

The next step is to calculate the matrix 
of classical overlaps $S$. 
To this end, we will need to compute the correlation functions 
of single-trace operators with  double-trace operators to 
$\cO (g_2) $. 
%gg
We will not  need the correlation functions of 
two different double-trace operators, because 
these overlaps are of 
 $\cO(g_2^2)$.
Hence, it is possible to treat each double-trace operator
independently 
and write the expressions \eqref{s},
\eqref{d} and \eqref{t} as two by two matrices.

Thus the classical overlap is given by
\bea 
S=\uno +g_2s \ ,
\eea
where
\bea \label{s}
s=\left (\begin {array}{clcr}
           &0     &C_{\rm free} \\
           &C_{\rm free}   &0  \\
          \end{array} \right ) \ ,
\eea   
and 
\bea
\label{103}
 C_{\rm free}=\sum_{\rm perm'}C_{\rm free}^{p} \ ,
\eea
where
\bea
 C_{\rm free}^{p}=\frac{(-1)^{p_2}}{\pi^{p_3}\sqrt{y^{p_1-1}(1-y)^{p_2-1}J}}
\prod_{a=1}^{p_1}\frac{\sin(\pi m_{p(a)} y)}
{m_{p(a)}-n_a/y}\prod_{b=1}^{p_2}\frac{\sin(\pi m_{p(b+p_1)} y)}
{m_{p(b+p_1)}-k_b/(1-y)} \ .
\eea
The sum in \eqref{103} is over all 
the admissible permutations of the ${\{m_{i}\}}$,
which label the barred BMN operator, 
as on the left hand side of \eqref{a+b+c}.
A permutation is admissible only when the permuted numbers belong to
$\phi$'s of the same flavour. 
 
Our next goal is to determine the anomalous dimension matrix $T$, 
\bea
T=d+g_2t \ ,
\eea
where the diagonal part $d$ contains the anomalous dimensions, as
in \eqref{matrix-d}, 
\bea \label{d}
d=\lambda'\left (\begin {array}{clcr}
           &\sum_{a=1}^{p_3}m_{a}^2/2     &0 \\ \cr
           &0            &\sum_{a=1}^{p_1} n_{a}^2/2y^2+
\sum_{a=1}^{p_2} k_{a}^2/2(1-y)^2  \\
          \end{array} \right )
\ , 
\eea  
and 
 \bea \label{t}
t=\lambda' \left (\begin {array}{clcr}
           &0        &t_{12} \\
           &t_{21}   &0  \\
           \end{array} \right )
\ . 
\eea   
$t_{12}$ can be read from \eqref{a+b+c}, and is given by
\be \label{t12defn}
t_{12}=\sum_{\rm perm'} C_{\rm free}^{p} (a+b+c) \ .
\ee
The coefficients $a$ and $c$ are given by the anomalous dimensions
of the first and second operators in the definition of $\cT$,
\be \label{acrest}
a =\sum_{a=1}^{p_1}\frac{n_{a}^2}{2y^2} \ , \qquad
c =\sum_{a=1}^{p_2}\frac{k_{a}^2}{2(1-y)^2} \ .
\ee
The contributions of 
$\sum_{\rm perm'} C_{\rm free}^{p}\,a$ and 
$\sum_{\rm perm'} C_{\rm free}^{p}\, c$ to
$t_{12}$ in \eqref{t12defn} factorise, to  give
\be 
C_{\rm free} \,a \ , \qquad C_{\rm free} \,c \ ,
\ee
respectively.

The remaining contribution to 
$t_{12}$ is $\sum_{\rm perm'} C_{\rm free}^{p}\,b$.
It can be extracted from the coefficient of the corresponding three-point
function following the same logic as in section 3.
These three-point functions of generic BMN operators with arbitrary numbers
of scalar impurities are computed in the following section.
Our result is
\bea
\label{112}
\sum_{\rm perm'} C_{\rm free}^{p}\,b =&&
\sum_{\rm perm'} C_{\rm free}^{p} \, \frac{1}{2}
\left( \sum_{a=1}^{p_1}m_{p(a)}\left(m_{p(a)}-\frac{n_{a}}{y}\right)
+\sum_{a=1}^{p_2} m_{p(a+p_1)}
\left(m_{p(a+p_1)}-\frac{k_{a}}{1-y}\right)\right. \nonumber\\
&&+\frac{1}{2}
\sum_{(a,b)}^{p_1}(m_{p(a)}-n_{a})(m_{p(b)}-n_{b})
+\frac{1}{2}\sum_{(a,b)}^{ p_2}(m_{p(p_1+a)}-k_{a})(m_{p(p_1+b)}-k_{b})
\nonumber\\
&&+\left. \frac{1}{2}\sum_{a=1}^{p_1}
\sum_{b=1}^{p_2}(m_{p(a)}-n_{a})(m_{p(p_1+b)}-k_{b})
\right) \ . 
\eea
%vv
The double sum summation notation
$(a,b)$ means that
we do not distinguish between the pair  
$a,b$ and the pair  $b,a$ ($a\neq b$).

As in section 3, the anomalous dimension matrix $\G$ 
in the isomorphic to string basis is given by
\bea
\G = d+ g_2 t' \ ,
\qquad t'=t-\frac{1}{2}\{s,d\} \ ,
\eea
where
\bea
\{s,d\}=\lambda' \left (\begin {array}{clcr}
           &0        &C_{\rm free}(\d_1+\d_2+\d_3) \\
           &C_{\rm free}(\d_1+\d_2+\d_3)       &0  \\
           \end{array} \right ) \ ,
\eea
and $\d_i$ is the anomalous dimension of the $i^{\rm th}$ operator
($i=1,2,3$). 
After some algebra,  
we obtain the final result:
\bea 
\label{field}
\G_{12}= &&\frac{\lambda'g_2}{4}\sum_{\rm perm'} C_{\rm free}^{p}
\left(\sum _{a=1}^{p_1}\left(m_{p(a)}-\frac{n_{a}}{y}\right)^2
+\sum_{a=1}^{p_2} \left(m_{p(a+p_1)}-\frac{k_{a}}{1-y}\right)^2 \right .
\nonumber\\
&&+ \left .  \sum_{(a,b)}^{p_1}(m_{p(a)}-n_{a})(m_{p(b)}-n_{b}) \right .
\\
&&+\left .\sum_{(a,b)}^{ p_2}(m_{p(p_1+a)}-k_{a})(m_{p(p_1+b)}-k_{b})
+\sum_{a=1}^{p_1}\sum_{b=1}^{p_2}(m_{p(a)}-n_{a})(m_{p(p_1+b)}-k_{b})\right )
%(\sum_{1\leq a\leq b\leq p_1}(m_{p(a)}-n_{a})(m_{p(b)}-n_{b})
%+\sum_{1\leq a\leq b\leq p_2}(m_{p(a)}-n_{a})(m_{p(b)}-n_{b})\nonumber\\
%+\sum_{1\leq a\leq p_1,1\leq b\leq p_2}(m_{p(a)}-n_{a})(m_{p(b)}-n_{b})) 
\ . 
\nonumber
\eea
This is our final expression for matrix elements in gauge theory.
In the next section we will compute the corresponding 
three-string amplitude and compare it to \eqref{field}. 
We will find perfect agreement.

%%%%%%%%%%%%%%%%%%%%%%%%%%%%%%%%%%%%%%%%%%%%%%%%%%%%%%%
\subsection{The results in string field theory }
In this subsection we assemble the basic ingredients of the SFT
calculation of the string amplitude for states with an arbitrary number
of scalar impurities.
The amplitude has the form \cite{SV1,SV2} 
(we refer the reader to Appendix A for more details)
\bea
\langle\Phi| {\sf P} \ket{V_B} \ , 
\eea
where $\langle\Phi|$ represents the three external string states, 
$\ket{V_B}$ is the kinematic part of the bosonic vertex \eqref{Vb}, 
and the prefactor {\sf P}  is given by 
\bea
{\sf P}=C_{\rm norm}\sum_{r=1}^{3}
\sum_{-\infty}^{\infty}\frac{\omega_{rn}}{ \mu \alpha_r}
\, \alpha^{r I \dagger}_{n}\alpha^{r I }_{-n} \ . 
\eea
$ C_{\rm norm}$ is defined in such a way that we get agreement with the
field theory calculation. Its value is taken to be
\bea \label{pre}
 C_{\rm norm}=-\frac{1}{2}g_2\frac{\sqrt{y(1-y)}}{\sqrt{J}}\, 
 (-1)^{\frac{1}{2} \sum_{\rr=1}^3\sum_{m=-\infty}^{+\infty}
\a^{\rr \, I\dag}_m \a^{\rr\, I}_m} \ .
\eea
The prefactor can act on the external bra--state and give
a sum of $2p_3$ terms, each of which has an external state identical  
to the initial  $\bra{\Phi}$, except one of the 
$\alpha_n$'s which has changed to $\alpha_{-n}$. Of course each of these
terms is multiplied by the corresponding $\omega_{rn} / { \mu \alpha_r}$. 
What we are left with is the action of the exponential in $\ket{V_B}$. 
In order to keep the comparison to field theory as simple as possible, 
we choose a certain set of associations
between the impurities of the third string and  
the impurities of the other
two strings. 
The final result will be a sum over all possible such associations,
i.e.~permutations of this set.

When an external oscillator 
has not been changed by the prefactor, 
the action of $ \ket{V_B}$ gives a factor of
$ \hat{N}_{n_a n_{a}'}^{3r}$ where $r=1,2,3$. But if the external oscillator 
has been changed by the prefactor, 
the action of $ \ket{V_B}$ gives a factor of
\beq
F^{3r}_{n_a -n_{a}'}= \hat{N}_{-n_a n_{a}'}^{3r} 
\frac{\omega_{3n_a}}{ \mu \alpha_3} 
+\hat{N}_{n_a -n_{a}'}^{3r} \frac{\omega_{rn_a'}}{ \mu \alpha_r}
=\hat{N}_{n_a -n_{a}'}^{3r}\left (\frac{\omega_{3n_a}}{\mu \alpha_3}
+\frac{\omega_{rn_a'}}{ \mu \alpha_r}\right)
\ . 
\eeq 
One can evaluate $F^{3r}_{n_a -n_{a}'}$ to get
\bea
F^{31}_{n_a -n_{a}'}&=&
(-1)^{n_a+ n_{a}'} 
\frac{\l'}{2 \pi \sqrt{y}}
\left( n_{a}-\frac{ n_{a}'}{y}\right)^2
\frac{\sin( \pi  n_{a} y)}{ n_{a}- n_{a}'/y} \ ,
\nonumber\\
F^{32}_{n_a -n_{a}'}&=&
(-1)^{n_a +1}\frac{\l'}{2 \pi  \sqrt{1-y}}( n_{a}-n_{a}'/(1-y))^2
\frac{\sin( \pi  n_{a} y)}{ n_{a}-
 n_{a}' / {(1-y)}} \ ,
\nonumber\\
F^{33}_{n_a -n_{a}'}&=&(-1)^{n_a+ n_{a}'+1}
\frac{2 \sin(\pi  n_{a} y)\sin(\pi  n_{a}' y) }
{\pi\mu  } \ ,
\nonumber\\
F^{11}_{n_a -n_{a}'}&=&\frac{2(-1)^{n_a +n_{a}'}}{4 \pi \mu  y} \ ,
\nonumber\\ 
F^{22}_{n_a -n_{a}'}&=&\frac{2}{4 \pi \mu (1-y)} \ .
\eea
We are now in position to write down the result for a given permutation.
This reads
\bea
\langle\Phi| {\sf P}\ket{V_B} =&&
C_{\rm norm}\left[ \sum_{a}^{p_1}F^{31}_{m_a -n_{a}}
\prod_{b\neq a}^{p_1}\hat{N}_{m_b n_{b}}^{31}
\prod_{b=1}^{p_2}\hat{N}_{m_{b+p_1} k_{b}}^{32}
+\sum_{a}^{p_2}F^{32}_{m_a -k_{a}}\prod_{b=1}^{p_1}\hat{N}_{m_b n_{b}}^{31}
\prod_{b\neq a}^{p_2}\hat{N}_{m_{b+p_1} k_{b}}^{32}
\right.
\nonumber\\
&&
+\left(\sum_{(a,b)}^{p_1}F^{33}_{m_a -m_{b}}\hat{N}_{n_a n_{b}}^{11}
+\sum_{(a,b)}^{p_1}F^{11}_{n_a -n_{b}}\hat{N}_{m_a m_{b}}^{33}\right)
\prod_{c\neq a,b}^{p_1}\hat{N}_{m_c n_{c}}^{31}
\prod_{c=1}^{p_2}\hat{N}_{m_{c+p_1} k_{c}}^{32}
\\
&&
+\left(
\sum_{(a,b)}^{p_2}F^{33}_{m_{p_1+a} -m_{p_1+b}}\hat{N}_{k_a k_{b}}^{22}
+\sum_{(a,b)}^{p_2}F^{22}_{k_a -k_{b}}\hat{N}_{m_{p_1+a} m_{p_1+b}}^{33}
\right)
\prod_{c=1}^{p_1}\hat{N}_{m_c n_{c}}^{31}
\prod_{c \neq a,b}^{p_2}\hat{N}_{m_{c+p_1} k_{c}}^{32}
\nonumber\\
&&
+
\left.
\left(\sum_{a=1}^{p_1}\sum_{b=1}^{p_2}F^{33}_{m_{a} -m_{p_1+b}}
\hat{N}_{m_{a} k_{b}}^{12}
+\sum_{a=1}^{p_1}\sum_{b=1}^{p_2}F^{12}_{n_{a} -k_{b}}
\hat{N}_{m_{a} m_{p_1+b}}^{33}\right)
\prod_{c \neq a}^{p_1}\hat{N}_{m_c n_{c}}^{31}
\prod_{c\neq b}^{p_2}\hat{N}_{m_{c+p_1} k_{c}}^{32}
\right] \ .
\nonumber
\eea
%vv
As before, in the double sum over 
all pairs $(a,b)$ which appears above 
we do not distinguish between the pair  $(a,b)$ and the pair  $(b,a)$.
Making use of the expressions for the Neumann matrices from
\cite{HSSV} it is now easy 
to obtain the final expression for 
the matrix element in string theory: 
\bea
\label{121}
\langle\Phi| {\sf P} \ket{V_B}=\frac{C_{\rm free}^{p}}{4}g_2 \l'
\bigg( \sum_{a=1}^{p_1}(m_{a}-\frac{n_{a}}{y})^2
+\sum_{a=1}^{p_2} (m_{a+p_1}-\frac{k_{a}}{1-y})^2\nonumber\\
+\sum_{(a,b)}^{p_1}(m_{a}-n_{a})(m_{b}-n_{b})+(m_{b}-n_{a})(m_{a}-n_{b})
\nonumber\\
+\sum_{(a,b)}^{p_2}(m_{p_1+a}-k_{a})(m_{p_1+b}-k_{b})+(m_{p_1+b}-k_{a})
(m_{p_1+a}-k_{b})
\nonumber\\
+\sum_{a=1}^{p_1}\sum_{b=1}^{p_2}(m_{a}-n_{a})(m_{p_1+b}-n_{b})+
(m_{p_1+b}-n_{a})(m_{a}-n_{b})
\bigg) \ .
\eea
One should note that in calculating the three string vertex we 
did not take into account terms where there were two contractions
of oscillators belonging to the same string if the prefactor had not
acted on one of these oscillators previously. This is so because these
terms are of order $(1 / \mu)^4={\l'}^2$, as can be easily seen.%
\footnote{
There is a subtlety in writing \eqref{121}, since 
the $C^p$ for each term in that equation are in fact different:
to each term in \eqref{121} one should associate 
the $C^p$ corresponding to the permutation of the indices which label
$m$, $n$ and $k$ appearing in the term considered.}

In order to get the final string theory result, we should not forget to sum
\eqref{121}
over all the admissible permutations, 
as we have done for the field theory result. 
%gg
This means that 
the first line of \eqref{121}
should be summed over all the possible 
permutations, while the remaining  lines 
should be summed over all 
permutations except those which exchange 
the $m$'s associated with the labels $a$ and $b$
(more precisely,
the permutations 
which exchange $m_a$ with $m_b$, 
$m_{p_1 + a}$ with $m_{p_1 + b}$, 
$m_a$ with $m_{p_1 + b}$ in the second, third and fourth  line
of \eqref{121}, respectively).
Including these permutations would result in a double-counting.
Once this sums are performed, 
we obtain perfect agreement with 
the field theory result \eqref{field}.

\section{A technical aside: calculation of general 
scalar BMN three-point functions}

This final section is devoted to the derivation of 
the expression \eqref{112}.

Our goal is to calculate
$\langle \cO_{\{n_{i}\}}^{J\cdot y} (x_1) \, 
\cO_{\{k_{i}\}}^{J\cdot (1-y)} (x_2) \, 
\bar{\cO}_{\{m_{i}\}}^{J} (0)\rangle$.
To simplify the notation, we will rename the operators in this 
Green's function as 
$\cO_{1}(x_1)$,  $\cO_{2}(x_2)$ and  $\bar\cO_{3}(0)$.
Let us assume that there are 
$f_{1}^{(i)}$ $\phi_{1}$'s, $ f_{2}^{(i)}$ $ \phi_{2}$'s,
$f_{3}^{(i)}$  $\phi_{3}$'s and $f_{4}^{(i)}$ $\phi_{4}$'s 
in the $i^{\rm th}$ operator
where $i=1,2,3$. 
We consider the case where $f_{1}^{(3)}=f_{1}^{(1)}+f_{1}^{(2)}$
with similar relations holding for the other three impurities.

There are of course many different diagrams. 
In order to deal efficiently with them,  let us
select a particular set of Wick contractions between the impurities 
of the barred operator and the impurities
of the unbarred  operators. 
Obviously, only impurities of the same flavour 
can be contracted. 
The full result will then 
be a sum over all the different permutations 
of such contractions. 

We start by considering the diagrams where the pure BMN part is taken 
in each of the three operators. These are drawn in Figure 7.
It is easy to see that there are 
$f_{1}^{(3)}!f_{2}^{(3)}!f_{3}^{(3)}!f_{4}^{(3)}!$ different 
contributions. Let us select one of them  
and draw the corresponding diagrams, Figure 7, in which 
the  $i^{\rm th}$ $\phi_{1}$ field of $\bar\cO_{3}(0)$ interacts with the 
 $n^{\rm th}$ $\phi_{1}$ field of $\cO_{1}(x_{1})$,  
while all  the other fields 
are freely contracted. 
The phase factor associated with  
these free contractions becomes, in the BMN limit:
\bea
P_{2}&=&\prod_{a \neq n}^{p_1}
\sum_{l_{a}=1}^{J_{1}} (\bar q_{a}r_{a})^{l_{a}}
\prod_{b=1}^{p_2}\sum_{l_{b}=
J_{1}+1}^{J} (\bar q_{p_1+b}p_{b})^{l_{b}}
\nonumber\\
&=&\prod_{a \neq n}^{p_1}\sum_{l_{a}}^{J_{1}}e^{-2 \pi i(m_{a}-\frac{n_{a}}
{y})\frac{l_{a}}{J}} 
\prod_{b=1}^{p_2}\sum_{l_{b}=J_{1}+1}^{J}e^{-2 \pi i(m_{p_1+b}
-\frac{k_{b}}{1-y})\frac{l_{b}}{J}}
\nonumber\\
&=&J^{p_3-1}\prod_{a \neq n}^{p_1}
\int_{0}^{y}dx e^{-2 \pi i(m_{a}-\frac{n_{a}}{y})x}
\prod_{b=1}^{p_2}\int_{y}^{1}dx e^{-2 \pi i(m_{p_1+b}-\frac{k_{b}  }{y})x}
\nonumber\\
&=& J^{p_3-1}\prod_{a \neq n}^{p_1} 
e^{-\pi i m_{a}y} \frac{\sin{\pi m_{a}y}}{( m_{a}-n_{a}/y)\pi}
\prod_{b=1}^{p_2} e^{-\pi i m_{p_1+b}y}
\frac{(-1)^{p_2}\sin{\pi m_{p_1+b}y}}{(m_{p_1+b}-k_{b}/(1-y))\pi} \ . 
\label{123}
\eea

Recall that \eqref{95}, \eqref{96} and  
\eqref{pure}, \eqref{comp} do not contain $q_1$. 
In \eqref{123} and in what follows $q_1$ is defined as
$q_1=\prod_{i=2}^{J}\bar{q_i}$,  
and 
$q_i= e^{2\pi i m_i / J}$
is the phase factor of the $i^{\rm th}$ impurity
($r_a$ and $p_a$ are the phase factors of $\cO_{1}$
and   $\cO_{2}$ respectively). 
\begin{figure} [ht]
\label{fig7}
\psfrag {fi} {\Huge${\phi_i}$}
\psfrag {zb} {\Huge${\bar{Z}}$}
\psfrag {dzb} {\Huge${\scriptstyle\overline{\partial_{\mu}Z}}$}
\psfrag {dfi} {\Huge${\scriptstyle{\partial_{\mu}\phi_i}}$}
\psfrag {z}{\Huge$Z$}
\psfrag {7a}{\Huge 7a}
\psfrag {7b}{\Huge 7b }
\begin{center}
{\scalebox{0.4}{
\includegraphics{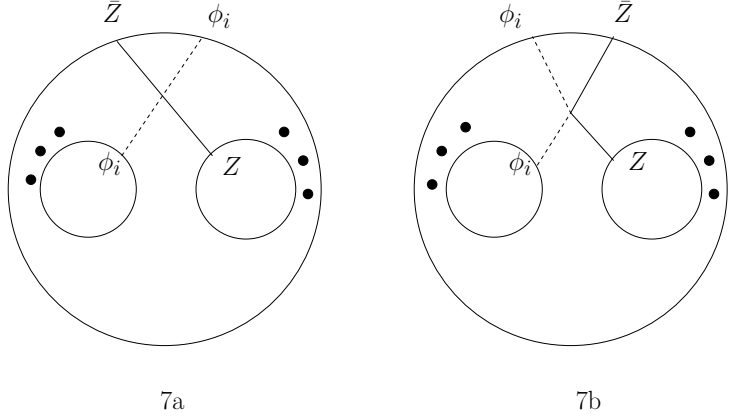}}
}
\end{center}
\caption{Feynman diagrams where a scalar impurity from 
$\cO_1$ interacts with a $Z$ field. The dots stand for 
impurities which have free contractions.
These diagrams are also accompanied by their mirror images, 
where the interaction occurs in the bottom part of the diagram.}
\end{figure}

Note that in obtaining the above formula we
have taken into account all the possible orderings 
of the freely contracting impurities.

We also need the phase factor associated with the 
fields which are involved in the interaction.
This is given by 
\bea
P_{1}= (\bar q_{i}^{J_{1}}-1)(\bar q_{i}-1)g^2
=\frac{-4\pi m_{i}}{J} e^{-\pi i m_{i}y}\sin{\pi m_{i}y}
\ . 
\eea
The total phase factor will be $P=P_{1}P_{2}$
\footnote{In this section the Lagrangian and Feynman rules
of \cite{CKT} are used.} 
Taking into account the normalisation of the operators, and
evaluating the space-time 
integral associated with the vertex (see \eqref{X1234}), 
one gets:
\bea
\label{125}
G_{3}^{(1)}=
\frac{J^{p_3}}{N\sqrt{J^{p_3-1}J_{1}^{p_1-1}J_{2}^{p_2-1}}}
\prod_{a \neq n}^{p_1}  \frac{\sin{\pi m_{a}y}}{m_{a}-n_{a} /y}
\prod_{b=1}^{p_2}
\frac{\sin{\pi m_{p_1+b}y}}{m_{p_1+b}-k_{b}  /(1-y)}(-1)^{p_2}
\nonumber\\
(-\l')\frac{m_{i}\sin{\pi m_{i}y}}{4 \pi^{p_3}}
\log{\frac{x_1^2x_2^2}{x_{12}^2}}  \ . 
\eea
In the previous expression we are keeping 
only the $\log x_{12}$ terms, which are relevant to determine
the coefficient $b$ in \eqref{a+b+c}, similarly to what 
we did in section 4.
We denoted by $G_{3}^{(1)}$ 
the part of the three-point function which corresponds
to the diagrams of Figure 7, 
and by $p_3=p_1+p_2$ the number of the impurities of $\cO_{3}$.

In order to make easier the comparison with the 
string theory result, 
we rewrite \eqref{125} as
\bea 
\label{dia1}
g_2C_{\rm free}^{p}b^{(1)}=
\frac{J^{p_3}}{N\sqrt{J^{p_3-1}J_{1}^{p_1-1}J_{2}^{p_2-1}}}
\prod_{a \neq n}^{k}  \frac{\sin{\pi m_{a}y}}{m_{a}-n_{a} /y}
\nonumber\\
\prod_{b=1}^{p_2}\frac{\sin{\pi m_{p_1+b}y}}{m_{p_1+b}-k_{b} /
(1-y)}(-1)^{p_2}
\frac{ m_{i}\sin{\pi m_{i}y}}{2  \pi^{p_3}} \ . 
\eea
Equation \eqref{dia1} corresponds to the first term in the first line 
of \eqref{112}. 
Until now we have considered only diagrams where the interacting
$\phi$ belongs to the operator
$\cO_1$. Of course, there are diagrams where it is an impurity in 
$\cO_2$ which interacts. 
These contributions produce the second term in the first line of 
\eqref{112}.

Now we consider the diagrams in Figure 8. 
\begin{figure} [ht]
\label{fig8}
\psfrag {fii} {\Huge ${\phi_i}$}
\psfrag {fij} {\Huge ${\phi_j}$}
\psfrag {zb} {\Huge ${\bar{Z}}$}
\psfrag {dzb} {\Huge ${\scriptstyle\overline{\partial_{\mu}Z}}$}
\psfrag {dfi} {\Huge ${\scriptstyle{\partial_{\mu}\phi_i}}$}
\psfrag {z}{\Huge $Z$}
\psfrag {8a}{\Huge 8a}
\psfrag {8b}{\Huge 8b }
\begin{center}
{\scalebox{0.4}{
\includegraphics{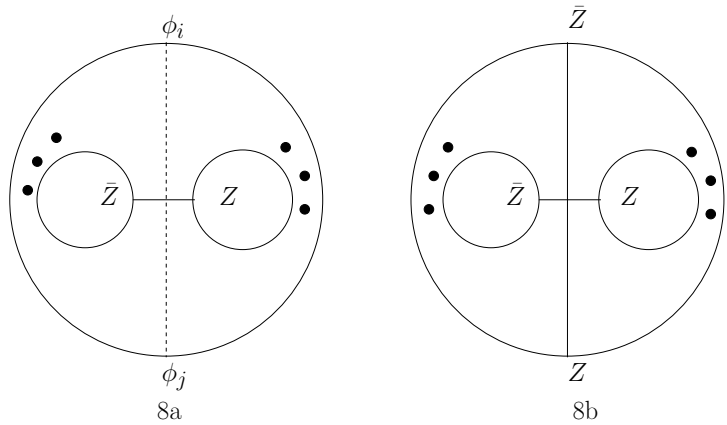}}
}
\end{center}
\caption{In Figure 8a
we take the pure BMN part in the external operator, whereas in 
8b we take the compensating term. In both cases we take 
the compensating term in $\cO_1$ and the pure BMN part in $\cO_2$.
Here $i$ and $j$ label the position of the corresponding 
$\phi$ in the barred BMN operator.}
\end{figure}
In these diagrams  we take for $\cO_1$ the compensating term,  
for  $\cO_2$ the pure BMN part, and for $\bar{\cO}_3$
the pure BMN part (Figure 8a) or  the compensating term
(Figure 8b).
In the case where two $\phi_1$'s interact,
the number of different diagrams is 
$f_{1}^{(3)}!f_{2}^{(3)}! f_{3}^{(3)}!f_{4}^{(3)}!
\frac{f_{1}^{(1)}(f_{1}^{(1)}-1)}{4}$.
This number is obtained as follows. There are 
$f_{1}^{(1)}(f_{1}^{(1)}-1) / {2}$ different ways to 
single out two $\phi_1$'s from the operator  $\cO_1$. 
This 
should be multiplied by the 
$f_{1}^{(3)}(f_{1}^{(3)}-1) / 2$ different ways in which 
we can  choose two $\phi_1$'s from the 
barred operator  $\bar {\cO}_3$, 
times the number 
$(f_{1}^{(3)}-2)!f_{2}^{(3)}! f_{3}^{(3)}!f_{4}^{(3)}!$ of 
independent free contractions of all the remaining 
impurities.

Evaluating the phase factor  
associated with the interacting fields we get
\bea
P_{1}= (\bar q_{i}^{J_{1}}-1)(\bar q_{j}^{J_{1}}-1)g^2.
\eea 
For the total phase factor one obtains
\bea 
P= J^{p_1+p_2-2}\prod_{a \neq i,j}^{p_1} e^{-\pi i m_{a}y} 
\frac{\sin{\pi m_{a}y}}{(m_{a}-n_{a} /y)\pi}
\prod_{b=1}^{p_2} e^{-\pi i m_{p_1+b}y}
\frac{\sin{\pi m_{p_1+b}y}}{(m_{p_1+b}-k_{b}  /(1-y))\pi}(-1)^{p_2}
\nonumber\\
g^2(-4)\sin{\pi m_{i}y}\sin{\pi m_{j}y} e^{-\pi i (m_{i}+m_{j})y}
\ . 
\eea 
The contribution to $G_3$ which corresponds 
to the diagrams of Figure 8 is therefore 
\bea
G_{3}^{(2)}=\frac{J^{p_3}}{N\sqrt{J^{p_3-1}J_{1}^{p_1-1}J_{2}^{p_2-1}}}
\prod_{a \neq i,j}^{p_1}  \frac{\sin{\pi m_{a}y}}{m_{a}-n_{a} /y}
\prod_{b=1}^{p_2}\frac{\sin{\pi m_{p_1+b}y}}{m_{p_1+b}-k_{b}  /(1-y)}(-1)^{p_2}
\nonumber\\
(-\l')\frac{\sin{\pi m_{i}y}\sin{\pi m_{j}y}}{4 \pi^{p_3 }}
\log{\frac{x_1^2x_2^2}{x_{12}^2}}
\ . 
\eea
{}From the last equation we can extract 
\bea 
\label{dia2}
g_2C_{\rm free}^{p}b^{(2)}=
\frac{J^{p_3}}{N\sqrt{J^{p_3-1}J_{1}^{p_1-1}J_{2}^{p_2-1}}}
\prod_{a \neq i,j}^{k}  \frac{\sin{\pi m_{a}y}}{m_{a}-n_{a} /y}
\nonumber\\
\prod_{b=1}^{p_2}\frac{\sin{\pi m_{p_1+b}y}}{m_{p_1+b}-k_{b}  /(1-y)}(-1)^{p_2}
\frac{\sin{\pi m_{i}y}\sin{\pi m_{i}y}}{2 \pi^{p_3}}
\ .
\eea
This term corresponds to the first term of the second line in 
\eqref{112}.
There are also diagrams where we consider the pure BMN part in 
$\cO_1$ and the compensating part in $\cO_2$ (rather than the opposite).
These terms produce the second term in the second line of \eqref{112}.

We now consider the last set of diagrams, which 
are represented in  Figure 9. 
\begin{figure} [ht]
\label{fig9}
\psfrag {fii} {\Huge${\phi_i}$}
\psfrag {fij} {\Huge${\phi_j}$}
\psfrag {fix} {\Huge${\phi_x}$}
\psfrag {fiz} {\Huge${\phi_z}$}
\psfrag {zb} {\Huge${\bar{Z}}$}
\psfrag {dzb} {\Huge${%\scriptstyle
\overline{\partial_{\mu}Z}}$}
\psfrag {dfi} {\Huge${\scriptstyle{\partial_{\mu}\phi_i}}$}
\psfrag {z}{\Huge$Z$}
\psfrag {9a}{\Huge 9a}
\psfrag {9b}{\Huge 9b }
\begin{center}
{\scalebox{0.4}{
\includegraphics{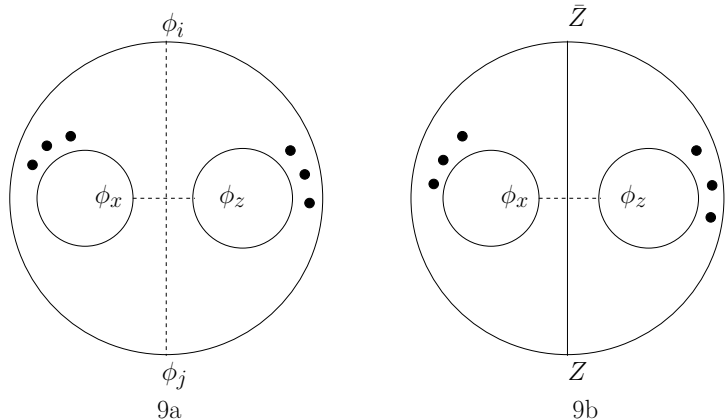}}
}
\end{center}
\caption{In Figure 9a we take the BMN part 
in the external operator, whereas
in 9b we take the compensating term. }
\end{figure}
In order to draw these diagrams we have used 
\eqref{pure}
for the operators $\cO_1$, $\cO_2$,  
and \eqref{pure}, \eqref{comp} 
for the barred operator.
In these diagrams  the 
$x^{\rm th}$ $\phi_1$ belongs to $\cO_1$, while the  $z^{\rm th}$ 
$\phi_1$ belongs to $\cO_2$.

In this case the phase factor  associated with the 
fields which interact becomes
\bea
P_{1}= (\bar q_{i}^{J_{1}}-1)(\bar q_{j}^{J_{1}}-1)(-g^2) \ . 
\eea 
For the total phase factor one obtains
\bea 
P= J^{p_3-2}\prod_{a \neq x}^{p_1} e^{-\pi i m_{a}y} 
\frac{\sin{\pi m_{a} y}}{m_{a}-n_{a} /y}
\prod_{b\neq z}^{p_2}\exp^{-\pi i m_{p_1+b}y}
\frac{\sin{\pi m_{p_1+b}y}}{m_{p_1+b}-k_{b}  /y}(-1)^{p_2-1}
\nonumber\\
\frac{(-g^2)(-4)}{\pi^{p_1-1}\pi^{p_2-1}}
\sin{\pi m_{i}y}\, \sin{\pi m_{j}y} e^{-\pi i (m_{i}+m_{j})y}
\ . 
\eea 
The contribution to $G_3$ which corresponds 
to the diagrams of Figure 9 is therefore 
\bea
G_{3}^{(3)}=\frac{J^{p-3}}{N\sqrt{J^{p_3-1}J_{1}^{p_1-1}J_{2}^{p_2-1}}}
\prod_{a \neq x}^{p_1}  \frac{\sin{\pi m_{a}y}}{m_{a}-n_{a} /y}
\prod_{b\neq z}^{p_2}\frac{\sin{\pi m_{p_1+b}y}}{m_{p_1+b}-k_{b} / 
(1-y)}(-1)^{p_2}
\nonumber\\
(-\l')\frac{\sin{\pi m_{i}y}\, \sin{\pi m_{j}y}}{4 \pi^{p_3 }}
\log{\frac{x_1^2x_2^2}{x_{12}^2}} 
\ . 
\eea
{}From this  equation we extract
\bea \label{dia3}
g_2C_{\rm free}^{p}b^{(3)}=
\frac{J^{p-3}}{N\sqrt{J^{p_3-1}J_{1}^{p_1-1}J_{2}^{p_2-1}}}
\prod_{a \neq x}^{p_1}  \frac{\sin{\pi m_{a}y}}{m_{a}-n_{a} /y}
\nonumber\\
\prod_{b\neq z}^{p_2}
\frac{\sin{\pi m_{p_1+b}y}}{m_{p_1+b}-k_{b}  /(1-y)}(-1)^{p_2}
\frac{\sin{\pi m_{i}y}\sin{\pi m_{i}y}}{2 \pi^{p_3}}
\ . 
\eea
Finally, by adding \eqref{dia1}, \eqref{dia2} and  \eqref{dia3}
(and the terms similar to  \eqref{dia1} and \eqref{dia2}, as discussed in 
the text), and summing over all permutations, 
it is immediate to obtain \eqref{112}.

%%%%%%%%%%%%%%%%%%%%%%%%%%%%%%%%%%%%%%%%%%%%%%%%%%%%%%%%%%%%%%%%%%%%%%%%%%

\section*{Acknowledgements} 

We would  like to thank Chong-Sun Chu 
for useful discussions. 
This work was partially supported by a PPARC SPG grant.
GG acknowledges a grant from the State Scholarship Foundation of
Greece (I.K.Y.).
  
%%%%%%%%%%%%%%%%%%%%%%%%%%%%%%%%%%%%%%%%%%%%%%%%%%%%%%%%%%%%

\newpage

\startappendix
\Appendix{the three-string vertex}

We first specify the notation and conventions 
used in pp-wave string field theory.
The combination $\a'p^+$ for the r-th string is denoted $\a_\rr$
and $\sum_{\rr=1}^3 \a_\rr =0$. As is standard in
the literature, we will choose a frame in which $\a_3=-1$
\be \label{frame}
\a_\rr = \a'p^+_{(\rr)} \, : \qquad \a_3=-1, \qquad \a_1=y, \qquad \a_2=1-y.
\ee
In terms of the $U(1)$ R-charges 
of the BMN operators in the gauge theory 
three-point function, $\langle \cO_1^{J_1}\,  \cO_2^{J_2} \, \bar\cO_3^{J}
\rangle$ 
we have
\be
y=\frac{J_1}{J}\ ,  \qquad 1-y=\frac{J_2}{J}, \qquad y \in (0, 1) \ , 
\ee
and $J=J_1+J_2$. 

The effective SYM coupling constant \eqref{lampr} in the frame \eqref{frame}
takes the simple form
\be \label{lamprsim}
 \l' =  \frac{1}{(\mu p^+ \a')^2}\ \equiv \frac{1}{(\mu \a_3)^2}\ 
= \frac{1}{\mu^2}.
\ee
Here $\mu$ is the mass parameter which appears in the pp-wave metric, in
the chosen frame it is dimensionless\footnote{It is $p^{+}\mu$ which is
invariant under longitudinal boosts and is frame-independent.} and the
expansion in powers of $1/\mu^2$ is equivalent to the perturbative
expansion in $\l'$. Finally,  the frequencies are defined via
\be
\omega_{\rr m}= \sqrt{m^2+(\mu\a_\rr)^2} \ .
\ee

The three-string vertex $\ket{H_3}$ 
can be represented as a ket-state in the tensor product of
three string  Fock spaces. It has the form \cite{SV1,SV2}
\be
\label{139}
{1\over \m} 
\ket{H_3} = {\sf P}  \ket{V_F} \ket{V_B}
\d \Big(\sum_{\rr=1}^{3} \a_\rr \Big)
\ ,
\ee
where the kets $\ket{V_B}$ and  $\ket{V_F}$ are constructed to satisfy
the bosonic and fermionic kinematic symmetries, and $\a_\rr$ are defined 
in \eqref{frame}.
The bosonic factor $\ket{V_B}$ is given by
\be 
\label{Vb}
\ket{V_B} = \exp\left( 
\frac{1}{2} \sum_{\rr,\rs=1}^3 \sum_{m,n=-\infty}^\infty
\sum_{I=1}^8 \a^{\rr\, I\dag}_{m} \Nh_{mn}^{\rr\rs}  \a^{\rs\,
I\dag}_{n}\right) \ket{0}  \ket{0}  \ket{0}   \ ,
\ee
where the $\Nh_{mn}^{\rr\rs}$ are the Neumann matrices in the
BMN-basis of string oscillators. 
The complete perturbative expansion of the Neumann matrices 
in the pp-wave background
in the vicinity of  $\mu=\infty$, 
was constructed in \cite{HSSV}%
\footnote{We refer the reader to the Appendix of 
Ref.~\cite{CK}
for some useful properties of the perturbative 
Neumann matrices and relations
between different string-oscillator bases.}.
The fermionic factor $\ket{V_F}$ 
is not going to be relevant for the present
paper, where only external bosonic string states are considered. 

The prefactor
${\sf P} $ is a polynomial in the bosonic and fermionic
oscillators, 
and is determined from imposing the remaining symmetries 
of the pp-wave background.  The relevant for us bosonic part of the prefactor,
as determined by Spradlin and Volovich in \cite{SV2}, reads
\bea \label{pref}
{\sf P}=C_{\rm norm}\sum_{r=1}^{3}\sum_{-\infty}^{\infty}
\frac{\omega_{rn}}{\mu \alpha_r}
\alpha^{r I \dagger}_{n}\alpha^{r J }_{-n} \, v_{IJ}\ , 
\eea
where $v_{IJ}={\rm diag}(\uno_4, -\uno_4)$, 
and the overall normalisation
$ C_{\rm norm}$ is left undetermined.
Notice that it is the expression for $v_{IJ}$ which leads 
to the relative minus sign
between the string amplitudes involving states with excitations along the
two different $SO(4)$'s as
e.g.~in \eqref{sv1} and \eqref{sv3}.

\Appendix{notation and conventions in gauge theory}
\label{appendixnc}
We write the bosonic part of the  $\N =4$ Lagrangian as
\be
{\cal{L}} = {2\over g^2} \ \Tr \left( {1 \over 4} F_{\mu \nu}F_{\mu \nu} + 
{1\over 2}(D_\mu \phi_i) (D_\mu \phi_i ) 
-{1\over 4} [\phi_i ,\phi_j][\phi_i ,\phi_j]\right) 
\ , 
\ee 
where $\phi_i$, $i=1, \ldots , 6$ are the six real scalar fields
transforming under an R-symmetry group $SO(6)$. 
The  covariant derivative is 
 $D_\mu \phi_i = \partial_{\mu}\phi_i  - i [A_\mu, \phi_i ]$, 
where 
$A_\mu = A_\mu^{a} T^a$, and $F_{\mu \nu} = \partial_{\mu}A_{\nu} - 
\partial_{\nu}A_{\mu} - i [ A_{\mu} , A_{\nu}]$.

If we define the complex combination
\beq
\label{compbas}
 Z  = {\phi_5 + i \phi_6 \over \sqrt{2}} \ , 
\eeq
the $\N =4$ Lagrangian can be re-expressed as 
\be
{\cal{L}} = {2\over g^2} \ \Tr \left( {1 \over 4} F_{\mu \nu}F_{\mu \nu} + 
\overline{(D^\mu Z )} (D_\mu Z  ) + {1\over 2} (D^\m \phi^i) (D_\m \phi^i)  
\right) 
+ V_F + V_D \ , 
\ee 
where the F-term and D-term potential are 
\beqa
\label{effe}
V_F & = &
% - {2\over g^2}\  \Tr \left( 
% [ \phi^I , \phi^J ] [\bar\phi_I ,  \bar\phi_J ] \right) 
% \ = \  
- {2\over g^2} \ \Tr \left(\, 2\, Z \phi_i \bar{Z} \phi_i - 
\phi_i \phi_i ( Z \bar{Z}+ \bar{Z} Z) + \cdots \right) \ , 
\\ 
V_D 
\label{di}
& = &  
% {1\over 2} {2\over g^2}\  \Tr \left( [ \phi^I , \bar\phi_I ] 
%  [ \phi^J , \bar\phi_J ] \right) \ =\ 
- {2\over g^2} \ \Tr \left(\,  ZZ \bar{Z}\bar{Z} - Z \bar{Z}Z \bar{Z}
+ \cdots \right) \ ,
\eeqa
are the F-term and D-term of the scalar potential respectively. 
In the last equalities the dots stand for impurity flavour changing terms, 
which mutually cancel between the F- and the D-term. 

Our $SU(N)$ generators are normalised as
\beq
\Tr \left( T^a T^b \right) = \delta^{ab} \ , 
\eeq
so that, for example,  
\beqa 
\left< Z^{i}_{j}(x) \bar{Z}^{l}_{m}(0) \right> = 
{g^2\over 2} \, \delta^{i}_{m} \delta^{l}_{j}\, \Delta (x) 
\ \ , \ \ \Delta (x)= {1\over 4\pi^2 x^2 } \  .
\eeqa

Finally, we will use the definitions $J:= J_1 + J_2$ and $J_1 = y\cdot J$, 
where $y\in (0, 1)$.

%%%%%%%%%%%%%%%%%%%%%%%%%%%%%%%%%%%%%%%%%%%%%%%%%%%
\Appendix{summing over BMN phase factors}
We report here the expressions for the coefficients 
$P_I $ and $P_{II}$ which arise after summing over 
the BMN phase factors in the interacting diagrams 
of section \ref{sec-mix}. Defining 
\beq
q = e^{2\pi i m /  J} \ , \qquad 
q_1 = e^{2\pi i n /  J_1} \ ,
\eeq
the expressions for $P_{I}$ and $P_{II}$ are given by 
\beq
P_{I} = \sum_{l=0}^{J_1}\  (\bar{q} q_1)^{l}
\ \bar{q} \ , \qquad 
P_{II} = \sum_{l=0}^{J_1} \ (\bar{q} q_1)^{l}
\ . 
\eeq
We also need to evaluate 
the quantity  $2(P_{I} + \bar{P}_{I}) - 2(P_{II} + \bar{P}_{II})$, 
which in the BMN limit is
\beqa
\label{totalpf}
2(P_{I} + \bar{P}_{I}) - 2(P_{II} + \bar{P}_{II}) 
%&=& 
%2\ \sum_{l=0}^{J_1} \ (\bar{q} q_1)^{l}\  (\bar{q} - 1 )
%\ + \  {\rm c.c.} 
%\nonumber \\ \ \cr
&=&
 - {8 m \over {m - n/  y}}\,  \sin^2 \pi m y
\ . 
\eeqa
%%%%%%%%%%%%%%%%%%%%%%%%%%%%%%%%%%%%%%%%%%%%%%%%%%
\Appendix{the function $X$}
The expression for three-point functions of BMN operators with 
scalar, vector, or mixed impurities involves the ubiquitous  
integral
\beq
\label{X1234}
 X_{1234} = \int d^4z \ 
\Delta (x_1 - z) \Delta (x_2 - z) \Delta (x_3 - z) \Delta (x_4 - z) 
\ \ .
\eeq
$X_{1234}$ develops  a $\log x_{12}^2$ term  $X$ 
as $x_1$ approaches $x_2$, which repeatedly appears 
in section \ref{sec-mix}. The expression for $X$ is 
\cite{CKTvec}
\beq
\label{X}
X\ := \ \left. X_{1234}\right|_{x_3 = x_4}\  =  \
%- {1 \over 2^8 \pi^6}  
% \left(  {1\over x_{13}^2 x_{23}^2} \right)\ ( x_{12}\Lambda)^2 = 
{\log \, ( x_{12}\Lambda)^{-1} \over 8 \pi^2 \, (4 \pi^2 x_{31}^2)^2}
\ .
\eeq
%%%%%%%%%%%%%%%%%%%%%%%%%%%%%%%%%%%%%%%%%%%%%%%%%%%%

%%%%%%%%%%%%%%%%%%%%%%%%%%%%%%%%%%%%%%%%%%%%%%%%%%

%%%%%%%%%%%%%%%%%%%%%%%%%%%%%%%%%%%%%%%%%%%%%%%%%%%%%%%%%%%

\end{document}